\documentclass[reprint,showpacs,aps,amsmath,amssymb,prd,nofootinbib,floatfix]{revtex4-1} 
\usepackage{graphicx}
\usepackage{dcolumn}
\usepackage{bm}
\usepackage{ulem} 
\usepackage{float}
\usepackage[usenames]{color}

\usepackage{times}
\usepackage{mathptmx}

\usepackage{amssymb,amsmath}
\usepackage{latexsym}
\usepackage{color}
\usepackage{cancel}
\newcommand{\beq}{\begin{equation}}
\newcommand{\eeq}{\end{equation}}
\newcommand{\beqn}{\begin{eqnarray}}
\newcommand{\eeqn}{\end{eqnarray}}
\newcommand{\pa}{\partial}

\newcommand{\varep}{\varepsilon}

\def\zeroD{\stackrel{(0)}{\mathstrut D}\hspace{-1.4mm}}
\def\zeroDelta{\stackrel{(0)}{\mathstrut \Delta}}
\def\zeroS{{d\hspace{-1.8mm}\stackrel{(0)}{\mathstrut S}\hspace{-1.mm}}}

\def\2pi{2\pi}

\begin{document}
\title{Coalescence of binary neutron stars in a scalar-tensor theory of
gravity}
\author{Masaru Shibata$^1$, Keisuke Taniguchi$^2$, Hirotada Okawa$^3$,
Alessandra Buonanno$^4$}
\affiliation{$^1$Yukawa Institute for Theoretical Physics, Kyoto University,
Kyoto 606-8502, Japan} 
\affiliation{$^2$Graduate School of Arts and Sciences, 
University of Tokyo, Komaba, Meguro, Tokyo 153-8902, Japan} 
\affiliation{$^3$CENTRA, Departamento de F{\'i}sica,
Instituto Superior T{\'e}cnico, Universidade T{\'e}cnica de Lisboa 
- UTL,Avenida Rovisco Pais 1, 1049 Lisboa, Portugal}
\affiliation{$^4$Maryland Center for Fundamental Physics and Joint 
Space-Science Institute, Department of Physics, University of
Maryland, College Park, MD 20742, USA}
\date{\today}

\begin{abstract}
We carry out numerical-relativity simulations of coalescing binary
neutron stars in a scalar-tensor theory that admits spontaneous
scalarization. We model neutron stars with realistic equations of
state. We choose the free parameters of the theory taking into account
the constraints imposed by the latest observations of neutron-star--
white-dwarf binaries with pulsar timing. We show that even within
those severe constraints, scalarization can still affect the evolution
of the binary neutron stars not only during the late inspiral, but
also during the merger stage. We also confirm that even when both
neutron stars have quite small scalar charge at large separations,
they can be strongly scalarized {\it dynamically} during the final
stages of the inspiral.  In particular, we identify the binary
parameters for which scalarization occurs either during the late
inspiral or only after the onset of the merger when a remnant,
supramassive or hypermassive neutron star is formed. We also discuss
how those results can impact the extraction of physical information on
gravitational waves once they are detected.
\end{abstract}

\maketitle

\section{Introduction}

Coalescing binary neutron stars are among the most promising sources
for the next-generation of kilometer-size gravitational-wave detectors
such as advanced LIGO, advanced Virgo, and KAGRA (or
LCGT)~\cite{LIGOVIRGO}. These detectors will be operational within the
next five years.  Based on the current estimates of event rates from
binary neutron stars, we expect that advanced detectors will observe
$\sim 0.4 \mbox{--} 400$ events per year at the distance of
$200$\,Mpc~\cite{RateLIGO,PlanLIGOVirgo}, which is the average
distance the advanced detectors will be sensitive to. Thus, likely,
the first detection(s) of gravitational waves will happen before the
end of this decade by observing gravitational waves emitted by
coalescing binary neutron stars.

One of the most interesting payoff of gravitational-wave observations
is the exploration of the validity of general relativity in the
strong-field dynamical regime. Scalar-tensor gravity is the simplest
and most well motivated class of alternative theories to general
relativity --- for example it has been postulated as a possible
low-energy limit of string theory. The most popular scalar-tensor
gravity theory was proposed by Jordan-Fierz-Brans-Dicke
(JFBD)~\cite{Jordan, Fierz, BD} (see also Ref.~\cite{will} for a
review). The JFBD theory depends on one single, constant parameter,
$\omega_{\rm BD}$, which determines the coupling strength between the
gravitational and scalar fields. This parameter has been strongly
constrained by a number of observations and
experiments~\cite{will2}. In particular, the experiments performed
using the Cassini spacecraft~\cite{Cassini} imply $\omega_{\rm BD}
\agt 4 \times 10^4$. 

There exist generalizations of the JFBD scalar-tensor theory that
still satisfy the weak but not the strong equivalence principle and
have richer phenomenology.  An interesting class of theories is the
one proposed by Damour and Esposito-Far\'ese~\cite{DEF1,DEF2,DEF3} 
in the early 90s.  In their models, $\omega_{\rm BD}$ is no
longer constant but depends on the scalar field $\phi$, i.e.,
$\omega(\phi)$. The latter can be chosen to be sufficiently large in
the weakly gravitating field of a star, such as the Sun, so that it
satisfies experimental tests~\cite{Cassini}, but it may be
significantly small, e.g., $\omega={\cal O}(1)$, in the strongly
gravitating field in the vicinity of massive neutron stars. Because
gravitational-wave observations will probe the strong-field dynamical
regime of coalescing neutron stars, they could detect or constrain
those alternative theories to general relativity.  Other modified
theories to general relativity have been proposed in the
literature~\cite{will2}. Among them, the $f(R)$ theories were
introduced as an alternative to the conventional dark-energy model, to
provide an explanation for the acceleration of the Universe. Those
theories can be recast into the form of a scalar-tensor
theory~\cite{SF}.  The Einstein-aether theory~\cite{JM} violates
Lorentz symmetry due to the existence of a preferred time direction at
each spacetime points. The free parameters in the Einstein-aether
theory have been constrained with a variety of
observations~\cite{phenomEA}.

Coalescing compact-body binaries offer a unique laboratory to test
alternative theories to general relativity through gravitational-wave
observations. To reach this goal, the two-body dynamics and
gravitational-wave emission in modified theories have been computed
analytically, in an approximated way, via the post-Newtonian
framework~\cite{E74,WZ,will,MW}, and more recently, also numerically,
solving the field equations with all the
nonlinearities~\cite{Healyetal,B2013,Bertietal}.  Here we focus on the
scalar-tensor theory by Damour and Esposito-Far\'ese
(DEF)~\cite{DEF1,DEF2,DEF3} and study its strong-field dynamical
regime by performing numerical-relativity simulations of coalescing
binary neutron stars.  As we shall see below, the possibility of
observing deviations from general relativity in the gravitational
waveforms in those theories may be possible because (i) neutron stars
in binaries can have large component masses (i.e., larger than the
canonical value $1.4 M_\odot$), (ii) the merger remnant is a neutron
star with a large mass, (iii) scalarization enhances the gravitational
interaction between the two neutron stars, reducing the time to
merger~\cite{B2013}, and (iv) neutron stars can be strongly scalarized
during the last stages of the inspiral and plunge even if one or both
neutron stars~\cite{B2013} had a quite small scalar charge at much
larger separations.  This latter phenomenon open the possibility of
observing deviations from general relativity via direct detection of
gravitational waves from binary neutron stars even if the indirect
observation of gravitational-waves via pulsar
timing~\cite{PSRJ1738,PSRJ0348} did not detect any deviation at much
larger separations.

Reference~\cite{B2013} has recently performed numerical-relativity
simulations of binary neutron stars in the DEF theory.  Here, we shall
investigate in more detail several interesting features found in
Ref.~\cite{B2013} and improve their work in different
directions. First, we start the numerical simulations from
quasiequilibrium configurations that consistently include also the
scalar field. By contrast, Ref.~\cite{B2013} set initially the scalar
field to zero. Second, Ref.~\cite{B2013} employed a simple polytropic
equation of state (EOS) with $\Gamma=2$ for the neutron star.  Whereas
this choice of the EOS may be acceptable for a qualitative study, it
does not describe very realistic neutron stars. As we shall see below,
the degree of scalarization in neutron stars does depend on the
EOS. Thus, if we want to make realistic predictions, we need to employ
realistic EOS, which is what we do here. Third, as a first study,
Ref.~\cite{B2013} focused only on the late stages of inspiral and
plunge. They did not investigate in any detail the merger phase. As we
shall find below, the frequency of gravitational waves emitted by the
newly-born, massive neutron star can be strongly modified due to
scalarization --- for example the frequency characteristics not only
depend on the EOS~\cite{BJ2012,hotoke2013b} but also on the scalar
field. Finally, an important difference between Ref.~\cite{B2013} and
our work is that we carry out the numerical simulation in the
so-called Jordan frame, while Ref.~\cite{B2013} employed the so-called
Einstein frame.

This paper is organized as follows. In Sec.~\ref{secII}, we describe
the basic equations of the scalar-tensor model employed here and the
numerical methods used to carry out the numerical simulations. We also
briefly discuss how we build the quasiequilibrium initial conditions
(for more details see Ref.~\cite{TSA13}).  In Sec.~\ref{sec:IV}, we
discuss the phenomenon of spontaneous scalarization for a single
neutron star and describe how we choose the free parameters in our
scalar-tensor model taking into account constraints from pulsar-timing
observations of binary pulsars~\cite{PSRJ1738,PSRJ0348}. In addition,
we explain how dynamical scalarization can occur in close binaries of
neutron stars.  In Sec.~\ref{sec:V}, we present the results of the
numerical simulations and discuss the effect of scalarization on the
gravitational waveforms during the last stages of inspiral, plunge, and
merger. Section~\ref{sec:VI} is devoted to a summary and a discussion
of future studies. Finally, in Appendix~\ref{appendixA} we check the
validity of the numerical code developed for scalar-tensor theories by
performing simulations of spherical neutron stars. In
Appendix~\ref{appendixC}, we study the numerical convergence of the
simulations and we estimate the numerical errors due to resolution.

Throughout this paper, we employ the geometrical units
$c=1=G$ where $c$ and $G$ are the speed of light and bare
gravitational constant, respectively.  Subscripts $a$, $b$, $c,
\cdots$ denote the spacetime components while $i$, $j$, $k$, and $l$
denote the spatial components, respectively.

\section{Numerical simulations in scalar-tensor gravity}
\label{secII}

\subsection{Basic equations}

We briefly summarize the basic equations of the JFBD-type
scalar-tensor theory in the 3+1 formulation. Scalar-tensor theories of
the simplest form are composed of the spacetime metric $g_{ab}$ and a
single real scalar field $\phi$ that determines the strength of the
coupling between the matter and the gravitational field. The action 
in the so-called Jordan frame is:
\beqn
S &=& \frac{1}{16 \pi G} \int d^4x \sqrt{-g}\,
\left[\phi {\cal R} - \frac{\omega(\phi)}{\phi}g^{ab} 
\nabla_a \phi\, \nabla_b \phi \right] \nonumber \\ 
&& -\int d^4x \sqrt{-g} \rho (1 + \varep) \,,
\eeqn
where ${\cal R}$ is the Ricci scalar associated with $g_{ab}$, 
$\rho$ is the rest-mass density, and $\varep$ is the 
specific internal energy. We note that in this paper, 
we describe the matter component with a perfect fluid. 
The equations of motion are 
\beqn
G_{ab}&=& 8\pi \phi^{-1} T_{ab} \nonumber \\
&+&\omega(\phi)\phi^{-2}
\biggl[(\nabla_a\phi) \nabla_b \phi-{1 \over 2}g_{ab}
(\nabla_c\phi)\nabla^c\phi \biggr] \nonumber \\
&&+\phi^{-1} (\nabla_a\nabla_b \phi - g_{ab} \Box_g \phi),\label{eq:JBD1} \\
\Box_g \phi &=& {1 \over 2\omega(\phi) +3} \Big[8\pi T - {d \omega \over d\phi}
(\nabla_c\phi)\nabla^c\phi\Big], \label{eq:JBD2} \\
\nabla_a T^a_{~b}&=&0,\label{eq:JBD3}
\eeqn
where $G_{ab}$ and $\nabla_a$ are the Einstein tensor and covariant
derivative associated with $g_{ab}$, $\Box_g$ is $\nabla_a \nabla^a$,
$\omega(\phi)$ determines the strength of the coupling between the
gravitational and scalar fields, and $T_{ab}$ is the stress-energy
tensor of the perfect fluid with $T=T_a^{~a}$. The matter is coupled
only to the gravitational field in the Jordan frame, as
Eq.~(\ref{eq:JBD3}) shows, and hence, the equations for the perfect
fluid are the same as those in general relativity in this frame. In
the following, we write Eqs.~(\ref{eq:JBD1}) and (\ref{eq:JBD2}) in
the 3+1 formulation.

The basic equations in the 3+1 formulation for the gravitational field
are derived simply by contracting $n^a n^b$, $n^a \gamma^b_{~i}$, and
$\gamma^a_{~i} \gamma^b_{~j}$ with Eq.~(\ref{eq:JBD1}).  Here,
$\gamma_{ab}$ denotes the spatial metric, and $n^a$ is the unit normal
to spatial hypersurfaces.  A straightforward calculation yields the
Hamiltonian constraint as
\beqn
R_k^{~k} + K^2 -K_{ij}K^{ij}&=&16\pi \phi^{-1} \rho_{\rm h} \nonumber \\
&+&\omega \phi^{-2} [\Pi^2 + (D_i\phi)D^i\phi] \nonumber \\
&+&2\phi^{-1}(-K \Pi + D_i D^i \phi), \label{eq:Ham}
\eeqn
where $R_k^{~k}$ is the three-dimensional Ricci scalar, $D_i$ the
covariant derivative with respect to the spatial metric, $\rho_{\rm
h}:=T_{ab}n^a n^b$, $\Pi:=-n^a \nabla_a \phi$, and $K_{ij}$ is the
extrinsic curvature with $K$ its trace. 

The momentum constraint is written as
\beqn
D_i K^i_{~j} - D_j K&=&8\pi \phi^{-1} J_j + \omega \phi^{-2} \Pi D_j\phi
\nonumber \\
&+&\phi^{-1}(D_j \Pi - K^i_{~j} D_i \phi),
\eeqn
where $J_i:=-T_{ab}n^a \gamma^b_{~i}$. 

Finally, the evolution equation is
\beqn
\pa_t K_{ij} &=&\alpha R_{ij}-8\pi \alpha \phi^{-1} 
\Big[S_{ij}-{1\over 2}\gamma_{ij}(S-\rho_{\rm h})\Big] \nonumber \\
&+&\alpha(-2K_{ik} K_j^{~k}+K K_{ij}) \nonumber \\
&-&D_i D_j \alpha +\beta^k D_k K_{ij}+K_{ik} D_j \beta^k+K_{kj} D_i \beta^k 
\nonumber \\
&-&\alpha \omega \phi^{-2} (D_i \phi) D_j \phi 
-\alpha \phi^{-1}
\Big[D_i D_j\phi-K_{ij}\Pi \nonumber \\
&+&{1 \over 2(2\omega+3)}\gamma_{ij}
\Big\{8\pi T + {d \omega \over d\phi}
(\Pi^2 -(D_k\phi) D^k\phi)\Big\}
\Big], \nonumber \\
\label{eq:JBD9}
\eeqn
where $R_{ij}$ is the spatial Ricci tensor and
$S_{ij}:=T_{ab}\gamma^a_{~i}\gamma^b_{~j}$ with $S$ its trace.
Equation~(\ref{eq:JBD9}) together with the Hamiltonian constraint
yields the following evolution equation for $K$:
\beqn
&&(\pa_t - \beta^k \pa_k) K 
= 4\pi \alpha \phi^{-1} (S+\rho_{\rm h})+\alpha K_{ij} K^{ij}-D_i D^i \alpha 
\nonumber \\
&&~~~~~~~+\alpha \omega \phi^{-2} \Pi^2 + \alpha \phi^{-1}
\Big[D_i D^i \phi - K \Pi \nonumber \\
&&~~~~~~~- {3 \over 2(2\omega+3)}
\Big\{8\pi T + {d \omega \over d\phi} (\Pi^2 -(D_k\phi) D^k\phi)\Big\}
\Big]. \label{eq:trKJBD}
\eeqn

The left-hand side of Eq.~(\ref{eq:JBD2}) is recast in the following
form
\beqn
\Box_g \phi 
= D_a D^a \phi + (D_a \ln \alpha) D^a \phi + (\nabla_a n^a) \Pi
+ n^a \pa_a \Pi,\nonumber \\
\label{eq:JBD4}
\eeqn
and then Eq.~(\ref{eq:JBD2}) is re-written into a set of equations
that are first order in the time derivatives 
\beqn
&&(\pa_t -\beta^k \pa_k)\phi = -\alpha \Pi,\label{eq:JBD5a} \\
&&(\pa_t -\beta^k \pa_k)\Pi = -\alpha D_i D^i \phi - (D_i \alpha)D^i\phi
+\alpha K \Pi \nonumber \\
&&\hskip 2cm +{\alpha \over 2\omega +3}
\Big[8\pi T - {d \omega \over d\phi}(\nabla_c\phi)\nabla^c\phi\Big].
\label{eq:JBD5}
\eeqn

The evolution equations for the gravitational fields are solved in the
Baumgarte-Shapiro-Shibata-Nakamura
formalism~\cite{shibatanakamura1995,baumgarteshapiro1998} with the
moving-puncture gauge~\cite{brandtbrugmann1997,clmz2006,bcckm2006} as
we have been doing in general relativity~\cite{sacra}. In particular,
we evolve the conformal factor $W := \gamma^{-1/6}$, 
the conformal metric $\tilde{\gamma}_{ij} := \gamma^{-1/3}
\gamma_{ij}$, the trace of the extrinsic curvature $K$, 
the conformally-weighted trace-free part of the extrinsic curvature
$\tilde{A}_{ij} := \gamma^{-1/3} ( K_{ij} - K
\gamma_{ij}/3)$, and the auxiliary variable $\tilde{\Gamma}^i := -
\partial_j \tilde{\gamma}^{ij}$. 
Introducing the auxiliary variable $B^i$ and a parameter $\eta_s$,
which we typically set to be $\sim m^{-1}$, $m$ being the total mass
of the system, we employ the moving-puncture gauge in the
form~\cite{bghhst2008}
\begin{eqnarray}
 ( \partial_t - \beta^j \partial_j ) \alpha &=& - 2 \alpha K , \\
 ( \partial_t - \beta^j \partial_j ) \beta^i &=& (3/4) B^i , \\
 ( \partial_t - \beta^j \partial_j ) B^i &=& ( \partial_t - \beta^j
  \partial_j ) \tilde{\Gamma}^i - \eta_s B^i .
\end{eqnarray}
The spatial derivative is evaluated by a fourth-order central finite
difference except for the advection terms, which are evaluated by a
fourth-order noncentered finite difference. We employ a fourth-order
Runge-Kutta method for the time evolution. For the scalar field, we
use the same scheme as those for the tensor field because the
structure of the equations is essentially the same. 

To solve the hydrodynamics equations, we evolve $\rho_* := \rho
\alpha u^t W^{-3}$, $\hat{u}_i := h u_i$, and $e_* := h \alpha
u^t - P / ( \rho \alpha u^t )$ with $u^a$, $P$, $h$ being the four
velocity, pressure, and specific enthalpy. The advection terms are
handled with a high-resolution central scheme by Kurganov and Tadmor
\cite{kurganovtadmor2000} with a third-order piecewise parabolic
interpolation for the cell reconstruction. For the EOS, we decompose the
pressure and specific internal energy into cold and thermal parts as
\begin{equation}
 P = P_{\rm cold} + P_{\rm th} \; , \; \varepsilon = \varepsilon_{\rm
  cold} + \varepsilon_{\rm th} .
\end{equation}
Here, $P_{\rm cold}$ and $\varep_{\rm cold}$ are functions of $\rho$,
and their forms are determined by nuclear-theory-based
zero-temperature EOSs.  Specifically, the cold parts of both variables
are determined using the piecewise polytropic EOS (see, e.g.,
Ref.~\cite{hotoke2013} for details).

Then the thermal part of the specific internal energy is defined from
$\varepsilon$ as $\varepsilon_{\rm th} := \varepsilon -
\varepsilon_{\rm cold}$. Because $\varepsilon_{\rm th}$ vanishes in
the absence of shock heating, $\varepsilon_{\rm th}$ is regarded as
the finite-temperature part. In this paper, we adopt a $\Gamma$-law
ideal gas EOS
\begin{equation}
 P_{\rm th} = ( \Gamma_{\rm th} - 1 ) \rho \varepsilon_{\rm th} ,
\end{equation}
to determine the thermal part of the pressure, and choose $\Gamma_{\rm
th}$ equal to 1.8 following \cite{BJO2010}. 


\subsection{Choice of the functional form of $\omega$ and equations for 
the scalar field} 
\label{sec:DEF}

To obtain a scalar-tensor model with spontaneous scalarization we use
the following function for $\omega(\phi)$
\beqn
{1 \over \omega(\phi)+3/2}= B\, \ln \phi, \label{eq:omega}
\eeqn
where $B$ is a free parameter. For reasons that will become clear
below we also introduce the field $\varphi$ defined as $\phi =
\exp(\varphi^2/2)$. If we want to compare our model (\ref{eq:omega})
with the one used in Refs.~\cite{DEF1,DEF2,DEF3}, we should consider
that Damour and Esposito-Far\'ese worked in the Einstein frame, while
we use the Jordan frame. In the Einstein frame one introduces the
field $\bar{\varphi}$~\footnote{We note that in
Refs.~\cite{DEF1,DEF2,DEF3} the authors denote the scalar field
$\bar{\varphi}$ with $\varphi$.}, which is related to $\phi$ through
the following equations
\beqn
\phi &=& \frac{1}{A^2(\bar{\varphi})}\,,\\
\alpha^2(\bar{\varphi}) &=& 
\left[\frac{\partial \ln A(\bar{\varphi})}{\partial \bar{\varphi}}\right]^2 = 
\frac{1}{2 \omega(\phi) + 3}\,,
\eeqn 
The simplest function that the authors of Refs.~\cite{DEF1,DEF2,DEF3} 
used to generate spontaneous scalarization is
\beqn
A(\bar{\varphi}) = e^{\frac{1}{2}\beta \bar{\varphi}^2}\,.
\eeqn
We have $\alpha_0= (\partial \ln A/\partial
\bar{\varphi})_{\bar{\varphi} = \bar{\varphi}_0}=\beta
\bar{\varphi}_0$ and $\beta_0= (\partial^2 \ln A/\partial
\bar{\varphi}^2)_{\bar{\varphi} = \bar{\varphi}_0}=\beta$. Moreover,
$\varphi = \sqrt{-2\beta}\bar{\varphi}$, so we find that $B =
-2\beta$. In summary, the parameters $(\bar{\varphi}_0,\beta_0)$ in
Refs.~\cite{DEF1,DEF2,DEF3} play a role similar to the parameters
$(\varphi_0,B)$ in this paper. As in previous 
works~\cite{DEF1,DEF2,DEF3,B2013}, we focus in this paper on the cases 
with $B \alt 10$. 

We note that when $\omega=$const, the scalar-field equation
(\ref{eq:JBD2}) is a simple wave equation for $\phi$, i.e., it is a
hyperbolic partial differential equation and it has a well-posed
initial value problem.  However, when $\omega$ is not a constant, such
as in Eq.~(\ref{eq:omega}), $\phi$ does not obey a wave equation
because of the presence of the second term in the right-hand side of
Eq.~(\ref{eq:JBD2}).  To derive a wave equation, at least in the far
zone, it is convenient to introduce $\varphi$ which is related to
$\phi$ by
\beqn
\phi=\exp(\varphi^2/2). \label{varphi}
\eeqn
Then, the equation for $\varphi$ reduces to
\beqn
\Box_g \varphi &=& 2\pi B T \varphi \exp(-\varphi^2/2) 
-\varphi (\nabla_c\varphi)\nabla^c\varphi. \label{eq:JBD2a}
\eeqn
In the far zone, the right-hand side of this equation falls off
sufficiently rapidly, and hence, $\varphi$ obeys a wave equation in
the far zone. 

We find it convenient to introduce a new variable 
$\Phi:=-n^a \nabla_a \varphi$ and replace 
Eqs.~(\ref{eq:JBD5a}) and (\ref{eq:JBD5}) by
\beqn
(\pa_t -\beta^k \pa_k)\varphi &=& -\alpha \Phi,
\\
(\pa_t -\beta^k \pa_k)\Phi &=& -\alpha D_i D^i \varphi 
- (D_i \alpha)D^i\varphi \nonumber \\
&& -\alpha \varphi (\nabla_a \varphi)\nabla^a
\varphi +\alpha K \Phi \nonumber \\
&&+ 2\pi \alpha B T \varphi \exp(-\varphi^2/2).\label{eq:varphi}
\eeqn
Here, the boundary condition for $r \rightarrow \infty$ should be
$\varphi=\varphi_0 \not= 0$ where $\phi_0=\exp(\varphi_0^2/2)$.  In
addition, we have $\Pi=\phi \varphi \Phi$ and $D_i \phi=\phi \varphi
D_i \varphi$, and, in a straightforward manner, we can replace $(\phi,
\Pi)$ to $(\varphi,\Phi)$ in all the gravitational-field equations.

Lastly, since in the far zone $\phi=e^{\varphi^2/2} \rightarrow 1 +
\varphi^2/2$, the asymptotic wave component of $\phi$ is $1 +
\varphi_0^2/2+\varphi_0 (\varphi-\varphi_0) +
O[(\varphi-\varphi_0)^2]$.  As we shall find in Sec.~\ref{sec:IV},
because of observational constraints $\varphi_0$ has to be
sufficiently small, and thus, the wave components in $\phi$ are also
quite small. This implies that {\em scalar-type gravitational waves},
which are directly related to $\phi$, are negligible in this theory,
although {\em scalar waves} associated with $\varphi$ are emitted to
carry energy and angular momentum from the system. 

\subsection{Equations of state employed}

In this paper, we employ APR4~\cite{APR4} and H4~\cite{H4} EOSs as in
Refs.~\cite{hotoke2013,hotoke2013b}. We remind that the APR4 EOS was
derived by a variational method with modern nuclear potentials for the
hypothetical components composed of neutrons, protons, electrons, and
muons. The H4 EOS was derived by a relativistic mean-field theory
including effects of hyperons. Here, for both EOSs, the maximum
allowed mass of spherical neutron stars is larger than $2M_{\odot}$
($\approx 2.20M_{\odot}$ for APR4 and $\approx 2.03M_{\odot}$ for H4),
and hence, the observational constraints by the latest discovery of
two-solar mass neutron stars~\cite{twosolar,PSRJ0348} are satisfied
for these EOSs.  The main difference between the two EOSs is that APR4
is a stiff but relatively soft EOS in which the stellar radius of a
spherical neutron star with canonical mass $1.35M_{\odot}$ is $\approx
11$\,km while H4 is a relatively stiff EOS in which the stellar radius
of a spherical neutron star with canonical mass $1.35M_{\odot}$ is
$\approx 13.5$\,km. This stiffness is quite important for determining
the properties of the scalarized neutron stars, as we shall describe
in Sec.~\ref{sec:IV}.

\subsection{Initial conditions for quasiequilibrium configurations}
\label{secIII}

We now explain how we prepare the initial conditions of the numerical
simulations using quasiequilibrium configurations for a binary in a
circular orbit with angular velocity $\Omega$.  To derive
quasiequilibrium configurations, we adopt the conformal flatness
formulation, that is
\beqn
\gamma_{ij}=\psi^4 f_{ij},\label{eq:conflat}
\eeqn
we assume the presence of a helical Killing vector, $(\pa_t + \Omega
\pa_\varphi)^a$, and the maximal slicing $K=0$~\cite{IWM}.
Here, $f_{ij}$ is the flat spatial metric.  For the fluid part, the
equations are the same as those in Einstein's gravity in the 
Jordan frame. Thus, assuming that the velocity field is
irrotational, the first integral of the hydrodynamics equations is
readily determined in the same manner as those in Einstein's
gravity~\cite{ST98}.

The basic equations for the tensor field are obtained from the
Hamiltonian and momentum constraints, together with
Eq.~(\ref{eq:trKJBD}) under the condition $K=0$. Except for the
modifications introduced by the presence of the scalar field $\phi$,
the equations are the same as that in Einstein's gravity. The
Hamiltonian and momentum constraints are, respectively, written as
\beqn
&&\zeroDelta \hspace{-1mm} \psi= -2\pi \phi^{-1} \rho_{\rm h} \psi^5 
-{1 \over 8} \tilde A_{ij} \tilde A^{ij} \psi^{5}\nonumber \\
&& \hskip 1cm
-{\psi^5 \over 8}\left[\omega \phi^{-2} \{\Pi^2 + (D_i\phi)D^i\phi\}
+2\phi^{-1} D_i D^i \phi\right], \nonumber \\
\label{eq:Ham1}
\eeqn
and
\beqn
\zeroD_i (\psi^6\tilde A^{i}_{~j})&=&\psi^6\Big[
8\pi \phi^{-1} J_j + \omega \phi^{-2}
\Pi \zeroD_j \phi  \nonumber \\
&&~~~+\phi^{-1}(\zeroD_j \Pi - \tilde A^i_{~j} \zeroD_i \phi)\Big],
\label{eq:Mom1}
\eeqn
where $\zeroDelta$ and $\zeroD_i$ are the Laplacian and
covariant derivative with respect to $f_{ij}$. $\tilde A_{ij}$ is the
tracefree conformal extrinsic curvature satisfying $K_i^{~j}=\tilde
A_i^{~j}$ for $K=0$ and its equation is derived from the evolution 
equation for $\gamma_{ij}$ with Eq.~(\ref{eq:conflat}) as
\beqn
\tilde A_{ij}={1 \over 2\alpha}
\left(f_{ik} \hspace{-1mm}\zeroD_j \beta^k+f_{jk} \hspace{-1mm}
\zeroD_i \beta^k
-{2 \over 3}f_{ij} \hspace{-1mm}\zeroD_k \beta^k\right), \label{eq:Aij}
\eeqn
where indices of $\tilde A_{ij}$, $\tilde A^{ij}$, and $\zeroD_i$
are raised and lowered by $f^{ij}$ and $f_{ij}$.  The condition $K=0$
yields
\beqn
\zeroDelta \hspace{-1mm}\chi &=& \chi \psi^4\biggl[
2\pi \phi^{-1} (2S+\rho_{\rm h})
+{7 \over 8}\tilde A_{ij} \tilde A^{ij}
\nonumber \\
&&~~~~~~+{1 \over 8}\omega \phi^{-2}\left\{
7\Pi^2 - (D_i \phi) D^i \phi \right\}\nonumber \\
&&~~~~~~+{3 \over 4\phi} 
\biggl\{ D_i D^i \phi - {2 \over (2\omega+3)} \nonumber \\
&&~~~\times
\Big(8\pi T + {d \omega \over d\phi} (\Pi^2 -(D_k\phi) D^k\phi)\Big)
\biggr\}\biggr], \label{eq:trKJBD1}
\eeqn
where $\chi:=\alpha \psi$. Note that we will replace the Laplacian
term of $D_i D^i\phi$ using the equation for $\phi$ (see below).

In addition to these equations, we have to solve the equation for
$\varphi$. If we simply impose that $\varphi$ satisfies 
the helical symmetry, we have
\beqn
\Phi=-\alpha^{-1}(\Omega \pa_\varphi + \beta^i \pa_i)\varphi. 
\eeqn
In this case, $D_i \phi$ and $\Pi$ in Eq.~(\ref{eq:JBD5}) behave as
$\propto r^{-1}$ in the far zone. If so, the spacetime cannot be
asymptotically flat because in the Hamiltonian constraint there exist
terms in the right-hand side that are proportional to $\Pi^2$ and
$(D_i\phi)D^i\phi$. Thus $\Pi$ and $D_i \phi$ have to be of order $r^{-2}$
in the far zone. To guarantee this condition, we simply set
$\Pi=0$. Then, Eq.~(\ref{eq:JBD5}) becomes an elliptic-type equation
so that $D_i \phi =O(r^{-2})$ is guaranteed in the far-zone. The boundary
condition to be imposed for $\varphi$ is $\varphi \rightarrow
\varphi_0$ for $r \rightarrow \infty$. Note that the resulting 
elliptic equation for $\varphi$ can be substituted in the right-hand
side of Eq.~(\ref{eq:Ham}). 

We compute the quasiequilibrium configurations using a new code which
is developed from a general-relativistic code originally implemented
in the spectral-method library LORENE~\cite{LORENE}.  We shall present
details of the numerical study of quasiequilibrium configurations in
Ref.~\cite{TSA13}.

\subsection{Definition of masses}

In scalar-tensor theories of gravity there are several definitions 
of masses. Here, we review them briefly. 

The ADM mass is defined as 
\beqn
M_{\rm ADM}:={1 \over 16\pi} \oint_{\infty} 
\gamma^{jk}\gamma^{il}(\pa_k \gamma_{ij}-\pa_i \gamma_{jk})\zeroS_l, 
\eeqn
where $\zeroS_l$ is the surface integral operator in flat space 
and $\oint_{\infty}$ denotes $\oint_{r \rightarrow \infty}$. 
In the conformally flat spatial hypersurface, the ADM mass may be defined as
\beqn
M_{\rm ADM}:=-{1 \over 2\pi} \oint_{\infty} 
Q \gamma^{jk} \pa_k \psi \zeroS_j,
\eeqn
where $Q$ is a function which reduces to unity when $r \rightarrow \infty$. 

From the asymptotic behavior of $\phi$ at $r \rightarrow \infty$, 
we can define the scalar mass $M_{\rm S}$ \cite{Lee74,will} as
\beqn
\phi =\phi_0 + {2M_{\rm S} \over r} + {\cal O}\left (\frac{1}{r^2}
\right ),
\label{eq:scalarcond}
\eeqn
where $\phi_0(=\exp(\varphi_0^2/2))$ is a constant close to unity
because $\varphi_0 \ll 1$ (see Sec.~\ref{sec:IV}).
Equation~(\ref{eq:scalarcond}) implies that the asymptotic behavior of
$\varphi$ is
\beq
\varphi= \varphi_0 + {M_{\varphi} \over r} + {\cal O}
\left (\frac{1}{r^2} \right ), \label{eq:varphiasy}
\eeq
where $M_{\varphi}$ is constant and related to $M_{\rm S}$ by $2M_{\rm
S}/\varphi_0$. In presence of a timelike Killing vector or helical
Killing vector, we can define the Komar mass~\cite{Komar}, which is
related to the ADM mass and the scalar mass by \cite{virial}
\beqn
M_{\rm K}=M_{\rm ADM} + 2M_{\rm S}. 
\eeqn

In addition, it is useful to define the tensor mass~\cite{Lee74}
\beqn
M_{\rm T}=M_{\rm ADM} + M_{\rm S},  
\eeqn
which, as Lee showed in Ref.~\cite{Lee74}, obeys a conservation law
similar to the one that the ADM mass obeys in general relativity.
Thus, in scalar-tensor theories of gravity, we find it more
appropriate to identify the neutron-star mass with the tensor mass
rather than the ADM mass. Henceforth, we shall use this identification
and set the neutron-star mass $M_{\rm NS}:= M_{\rm T}$.

\subsection{Simulation set-up and validation}

We perform numerical simulations using an adaptive-mesh refinement
code {\tt SACRA-ST} that was implemented by modifying the original
code for general relativity~\cite{sacra}.  As done for the simulations in 
Refs.~\cite{hotoke2013,hotoke2013b}, the semi-major
diameter of neutron stars is initially covered by $\approx 100$ grid
points (we refer to this grid resolution as high resolution).  For
APR4 and H4, the finest grid resolution is $\approx 0.17$ and
0.22\,km, respectively. We also perform lower-resolution simulations
covering the semimajor diameter by $\approx 67$ and 80 grid points
(we refer to these grid resolutions as low and medium resolutions),
and check that we achieve sufficient convergence to trust  
the conclusions of this paper (see Appendix B for details). 

We also confirm the validity of our code by performing (i) simulations
of spherical stars, (ii) longterm evolutions of scalarized spherical
neutron stars, and (iii) collapses of a scalarized neutron star to a
black hole.  The success of these tests give us confidence in our new
scalar-tensor code (see Appendix A for details). 

\section{Parameters choice for spontaneous scalarization in 
binary neutron stars}
\label{sec:IV}

\begin{figure*}[t]
\includegraphics[width=86mm,clip]{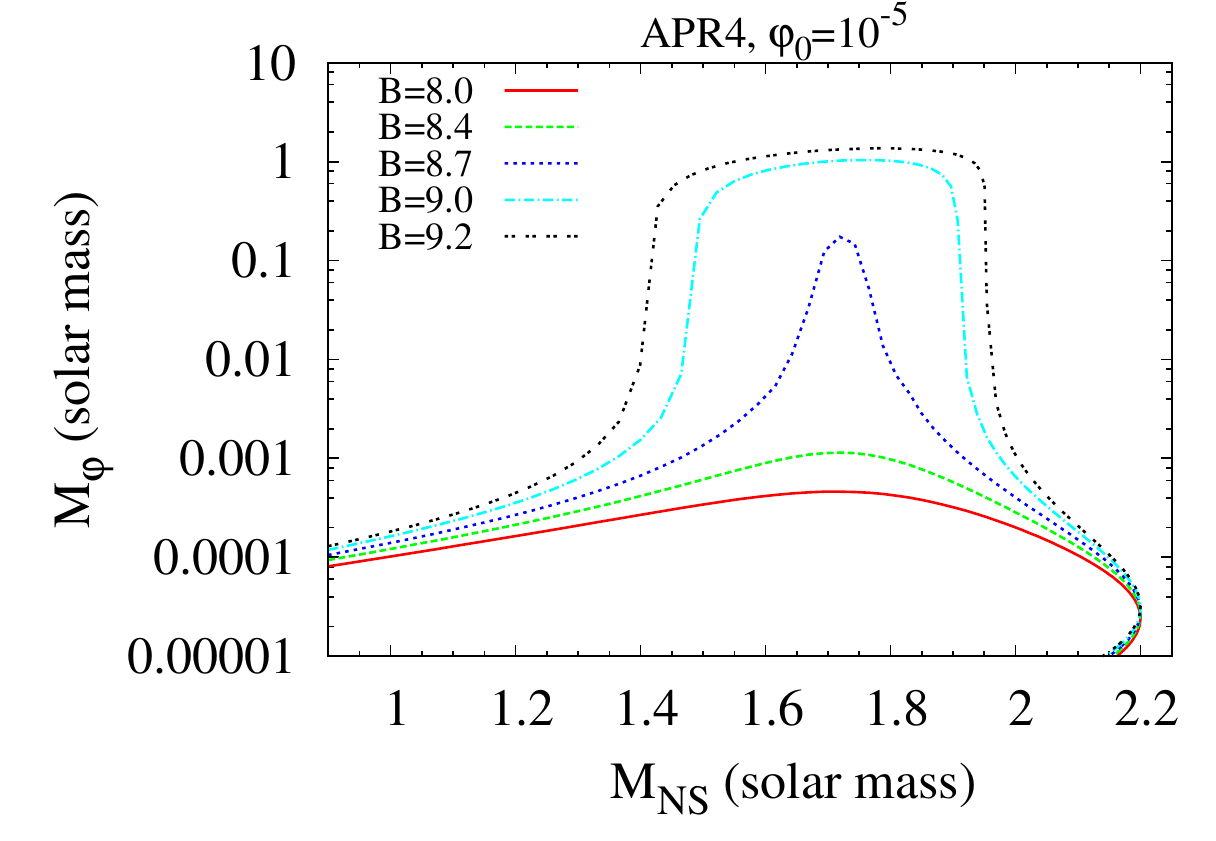}
\includegraphics[width=86mm,clip]{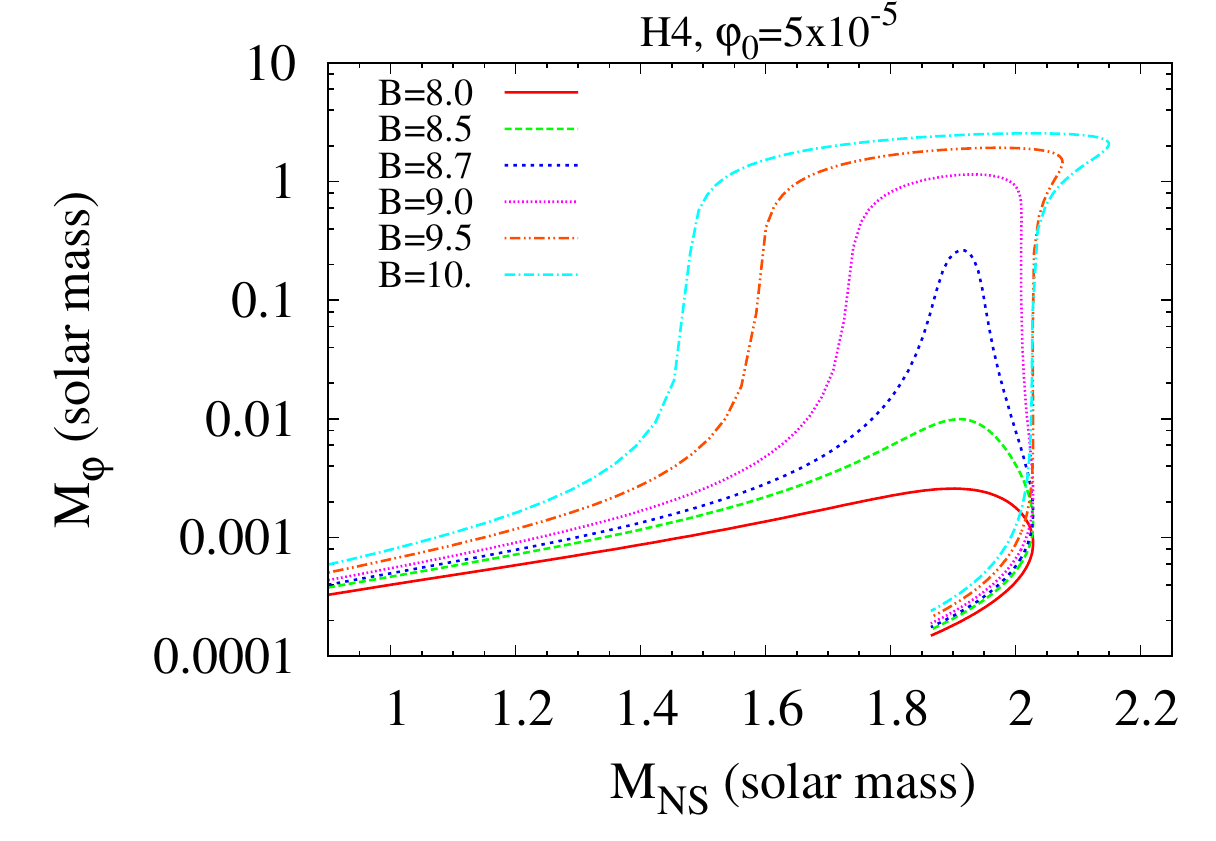}
\caption{We plot the value of $M_{\varphi}$ as a function of 
the neutron-star mass for spherical, isolated neutron stars using the
APR4 and H4 EOSs, and several values of $B$. The values of $\varphi_0$
are chosen to be $10^{-5}$ and $5 \times 10^{-5}$ for APR4 and H4
EOSs, respectively.}
\label{fig:MS}
\end{figure*}

In this section, we first review the key mechanism responsible for 
spontaneous scalarization in a single star and then present a physical
argument to explain why scalarization can occur in binary
systems even if the scalar charge at large separations were 
very small. Furthermore, for the EOSs employed in this paper, we
determine the values of $B$ and $\varphi_0$ such that they
satisfy the constraints imposed by pulsar-timing
observations~\cite{PSRJ1738,PSRJ0348}.  We shall perform numerical
simulations for those choices of the parameters.

\subsection{Spontaneous scalarization in an isolated star}
\label{appendixB}

Here, we follow Ref.~\cite{DEF1} and review the key idea 
underlying spontaneous scalarization. For simplicity we restrict 
the discussion to the static case and we neglect the gravitational 
field and nonlinear terms in $\varphi$. Within these approximations 
Eq.~(\ref{eq:JBD2a}) can be written as
\beqn
\Delta \varphi=2\pi B T \varphi, \label{eq:varphi0}
\eeqn
where $\Delta$ denotes the flat Laplacian. Assuming that relativistic
corrections are small, we have $T \approx -\rho <0$.  We also assume
that $T = {\rm const.}$, $B>0$, and set $k^2=-B T$.  Considering that
the star is spherically symmetric in isolation, we find that the
solution of Eq.~(\ref{eq:varphi0}) is~\cite{DEF1}
\beqn
\varphi=
\left\{
\begin{array}{ll}
\displaystyle {\cal A} {\sin(kr) \over r}    &~~ r \leq R,\\
\displaystyle {M_\varphi \over r}+\varphi_0     &~~ r \geq R,
 \end{array}
 \right.
\eeqn
where ${\cal A}$ is a constant and $R$ denotes the stellar radius. 
The continuity conditions of $\varphi$ and $d\varphi/dr$ at 
$r=R$ then yield
\beqn
&&{\cal A}={\varphi_0 \over k\cos(kR)},\\ 
&&M_\varphi=\varphi_0 [k^{-1}
\tan(kR)-R],
\label{eq:SS}
\eeqn
This suggests that for $kR \rightarrow \pi/2$, $\varphi$, as well as
$M_\varphi$, significantly increases, i.e., the scalarization occurs,
irrespective of the value of $\varphi_0$. Thus, the onset of
scalarization depends on three parameters, $B$, $T$, and
$R$. Then, if we assume $T \sim -\rho$ and use $\rho R^3 \sim
M_{\rm NS}$ where $M_{\rm NS}$ is the mass of the neutron star, we
have that $kR$ is proportional to $B^{1/2}(M_{\rm
NS}/R)^{1/2}$. Thus we conclude that the scalarization is determined
by two parameters: $B$ and the stellar compactness (or the
mass of the neutron star).

\begin{table*}[t]
\caption{The value of $F$ for the APR4 EOS with $\varphi_0 = 10^{-5}$ and
the H4 EOSs with $\varphi_0 = 5\times 10^{-5}$.  The unit of $F$ is
$M_\odot$. When ``---'' appears, it means that for such a model, the
relation~(\ref{eq:MF}) breaks down, thus scalarization occurs.  }
{\begin{tabular}{c|ccccc} \hline 
APR4 & &&$B$&& \\ $M_{\rm NS} (M_{\odot})$ 
& ~~~$8.0$~~~ & ~~~$8.5$~~~ & ~~~$9.0$~~~ & $9.5$ &
$10.0$ \\ \hline 
1.30 & 21 & 32 & 62 & $3.8 \times 10^2$ & --- \\ 
1.35 & 24 & 39 & 91 & --- & --- \\
1.40 & 27 & 48 & $1.6 \times 10^2$ & --- & --- \\ 
1.45 & 30 & 59 & $4.0 \times 10^2$ & --- & --- \\ 
1.50 & 34 & 75 & --- & --- & --- \\ \hline
\end{tabular}
}
\hskip 5mm 
{\begin{tabular}{c|ccccc} \hline 
H4 & &&$B$&& \\
$M_{\rm NS} (M_{\odot})$ & ~~~$8.0$~~~ & ~~~$8.5$~~~ & ~~~$9.0$~~~ & $9.5$ &
$10.0$ \\ \hline
1.30 & 14 & 18 & 24 & 34 & 54 \\
1.35 & 16 & 21 & 28 & 42 & 77 \\
1.40 & 17 & 23 & 34 & 55 & $1.3 \times 10^2$ \\
1.45 & 19 & 27 & 41 & 77 & $3.7 \times 10^2$ \\
1.50 & 22 & 31 & 51 & $1.2 \times 10^2$ & --- \\
\hline
\end{tabular}
}
\label{table:C}
\end{table*}

For $B < 0$ or $T > 0$ (i.e., for $BT>0$), the
solution of Eq.~(\ref{eq:varphi0}) in spherical symmetry is~\cite{DEF1} 
\beqn
\label{noscal}
\varphi=
\left\{
\begin{array}{ll}
\displaystyle {\cal A} {\sinh(kr) \over r}    &~~ r \leq R,\\
\displaystyle {M_\varphi \over r}+\varphi_0     &~~ r \geq R,
 \end{array}
 \right.
\eeqn
and the continuity conditions yield
\beqn
&&{\cal A}={\varphi_0 \over k\cosh(kR)},\\ 
&&M_\varphi=\varphi_0 [k^{-1}\tanh(kR)-R]. 
\eeqn
Here we set $k^2=B T$.  Thus, in this case, the scalarization is not
likely to occur for any value of $B$, $T$, and $R$.  This suggests
that for the ultra-relativistic case with $T=-\rho h +4P>0$ (and
$B>0$), the scalarization does not occur.

The above analysis suggests that when the scalarization {\it does not} 
occur, $M_\varphi$ is proportional to $\varphi_0$ and we can write
\beqn 
M_{\varphi}=F(M_{\rm NS},B) \varphi_0, \label{eq:MF}
\eeqn
where $F$ is a function that depends on $M_{\rm NS}$ and $B$. 
From a numerical analysis of spherical neutron stars in
equilibrium, we indeed find that this relation is satisfied as long as
the spontaneous scalarization does not set in.

In Fig.~\ref{fig:MS} we plot $M_\varphi$ as a function of the
neutron-star mass $M_{\rm NS}$ for the APR4 and H4 EOSs, using
$\varphi_0=10^{-5}$ and $=5 \times 10^{-5}$, respectively. We observe
the following interesting properties. If $M_{\rm NS}$ is smaller than
a critical value $M_{\rm NSc1}$, $M_\varphi$ is much smaller than
$M_{\rm NS}$.  The critical value depends strongly on the value of
$B$.  For larger values of $B$, $M_{\rm NSc1}$ is smaller, and hence,
spontaneous scalarization sets in for smaller neutron-star masses.  By
contrast, if $M_{\rm NS}$ is {\rm larger} than a critical value
$M_{\rm NSc2}$, $M_\varphi$ is again much smaller than $M_{\rm NS}$.
Thus, when neutron stars have sufficiently large masses spontaneous
scalarization never sets in.  This is due to the fact that for those
large masses, the relativistic effects are so significant that
$T=-\rho h+4P$ could be positive. This would imply that observations
of neutron stars with large masses, e.g., $\approx 2M_\odot$, may not
be very useful in constraining the value of $B$. Finally, for $M_{\rm
NSc1} < M_{\rm NS} < M_{\rm NSc2}$, neutron stars are spontaneously
scalarized, for certain values of $B$, e.g., $B \agt 8.5$ for the APR4
EOS.  Indeed, in these cases, $M_{\varphi}$ is on the order of $M_{\rm
NS}$. Using the qualitative analysis worked out at the beginning of
this section, in particular Eq.~(\ref{eq:SS}), we find that the value
of $M_\varphi$ could diverge when spontaneous scalarization occurs.
However, when using the realistic {\em nonlinear} equation for
$\varphi$, instead of Eq.~(\ref{eq:varphi0}), we find that nonlinear
effects always constrain $M_\varphi$ to be at most equal to the
neutron-star mass $M_{\rm NS}$.

\subsection{Condition for scalarization in inspiraling binary neutron stars}
\label{condition}

As described in the previous section, for an isolated, spherical
neutron star in which the scalarization has not occurred, the profile
of $\varphi$ is approximately described by Eq.~(\ref{eq:varphiasy}),
where $M_\varphi \ll M_{\rm NS}$. Given this field configuration, we
now suppose that the neutron star is in a binary system and it is not
yet spontaneously scalarized.  In this case it is natural to assume
that Eq.~(\ref{eq:MF}) gets approximately modified by the companion
star as
\beqn
M_{\varphi} \approx F(M_{\rm NS},B) \left(\varphi_0 +
{M_{\varphi} \over a}\right), \label{eq:MF2}
\eeqn
where $a$ is the orbital separation.  Namely, the value of $\varphi$
just outside the neutron star is enhanced by the presence of the
companion. (Note that for simplicity we are considering an equal-mass
(or nearly equal-mass) binary.) Solving Eq.~(\ref{eq:MF2}) for
$M_\varphi$ yields
\beqn
M_{\varphi} \approx F(M_{\rm NS},B) \varphi_0 
\left(1-{F(M_{\rm NS},B) \over a}\right)^{-1}, \label{eq:MF3}
\eeqn
and hence, $M_{\varphi}$ can increase steeply and can become on the
order of $M_{\rm NS}$ when $a \sim F(M_{\rm NS},B)$. Thus, even if the
values of $\varphi_0$ and $B$ are such that spontaneous scalarization
of the isolated neutron star is absent or it occurs only weakly, the
neutron star can be strongly scalarized if it is part of a binary
system and if the condition $a \alt F(M_{\rm NS},B)$ is
satisfied. Because this scalarization sets in when the neutron star is
part of a binary system, we denote it {\it dynamical} scalarization to
distinguish it from spontaneous scalarization.~\footnote{We note that
Ref.~\cite{B2013} simulated a binary configuration in which neutron
stars are not initially spontaneously scalarized and found that {\it
induced} scalarization can set in in the late inspiral. They also gave
a qualitative explanation of this phenomenon resorting to
energetically favoured arguments discussed in Ref.~\cite{GEF}.} This
property is indeed confirmed in our accompanying
paper~\cite{TSA13}. Let us now investigate when the condition $a \alt
F(M_{\rm NS},B)$ holds.

We list in Table~\ref{table:C} the values of $F$ for different
neutron-star masses and different values of $B$, for the two EOSs that
we use in this paper, notably APR4 and H4. (When ``---'' appears, it
means that for such a model the spontaneous scalarization does occur,
and thus, Eq.~(\ref{eq:MF}) no longer holds.) For binary neutron
stars, the merger occurs typically at $a=30$\,--\,45\,km $\approx
20$\,--\,$30M_\odot$ depending on the EOS. This implies that if $F$ is
smaller than $20$\,--\,$30M_\odot$, dynamical scalarization does not
occur during the inspiral stage. We find that for dynamical
scalarization to occur, $F$ has to be larger than at least
$20M_{\odot}$ for APR4 and $\sim 25M_{\odot}$ for H4.  As we see in
Table~\ref{table:C}, for $M_{\rm NS}=1.35M_{\odot}$, dynamical
scalarization can always set in before merger for APR4 EOS when $B
\agt 8.0$.  By contrast, for H4 EOS, dynamical scalarization can take
place only when $B\agt 9.0$ for $M_{\rm NS}=1.35M_{\odot}$.  These
properties are confirmed in our accompanying paper~\cite{TSA13}. 

Before ending this section, we present the analysis for unequal-mass
binary systems. Let $M_{\varphi,1}$ and $M_{\varphi,2}$ be the values
of $M_\varphi$ for stars 1 and 2.  Then, Eq.~(\ref{eq:MF2}) can be
rewritten in two equations
\beqn
M_{\varphi,1} &\approx& F_1 \left(\varphi_0 +
{M_{\varphi,2} \over a}\right), \label{eq:MF31} \\
M_{\varphi,2} &\approx& F_2 \left(\varphi_0 +
{M_{\varphi,1} \over a}\right), \label{eq:MF32}
\eeqn
where $F_1:=F(M_{\rm NS,1},B)$ and $F_2:=F(M_{\rm NS,2},B)$ with
$M_{{\rm NS},i}$ being the mass of neutron star $i$.
Equations~(\ref{eq:MF31}) and (\ref{eq:MF32}) yield
\beqn
M_{\varphi,1} &\approx& \varphi_0 F_1  
\left(1+{F_2 \over a}\right)
\left(1-{F_1 F_2 \over a^2}\right)^{-1},\\
M_{\varphi,2} &\approx& \varphi_0 F_2  
\left(1+{F_1 \over a}\right)
\left(1-{F_1 F_2 \over a^2}\right)^{-1}.
\eeqn
Thus, we expect the scalarization to occur when  
$a \approx \sqrt{F_1F_2}$ for both neutron stars  
approximately simultaneously. 

\subsection{Constraints from pulsar binary systems} \label{sec:IVA}

Pulsar timing observations of binary systems composed of a neutron
star and a white dwarf~\cite{PSRJ1738,PSRJ0348} impose the strongest
constraints on $B$ and $\varphi_0$ for a high value of $B \agt 5$. The
constraints come primarily from the fact that the scalar-wave
luminosity has to be substantially smaller than the gravitational-wave
luminosity.

The neutron-star masses measured in Refs.~\cite{PSRJ1738,PSRJ0348} are
$M_{\rm NS}=1.46^{+0.06}_{-0.05}M_{\odot}$ and $M_{\rm NS}=2.01 \pm
0.04 M_{\odot}$ at one--$\sigma$ error, respectively. As we shall find
below, those observations imply that neutron stars with masses $\alt
1.46M_{\odot}$ and $\agt 2.01M_{\odot}$ cannot be scalarized and that
the possible values of $B$, which depend on the EOS, are strongly
limited. Although Refs.~\cite{PSRJ1738,PSRJ0348} has already
constrained the DEF scalar-tensor model, they did it employing one
specific EOS for the nuclear matter~\cite{DEF3}.  As we have
emphasized when discussing Fig.~\ref{fig:MS}, the constraint on $B$
depends on the EOS. Therefore, our analysis, although similar to and
simpler than the one of Refs.~\cite{PSRJ1738,PSRJ0348}, pays special
attention to the dependence of the constraints on the EOS.

\begin{figure}[t]
\includegraphics[width=86mm,clip]{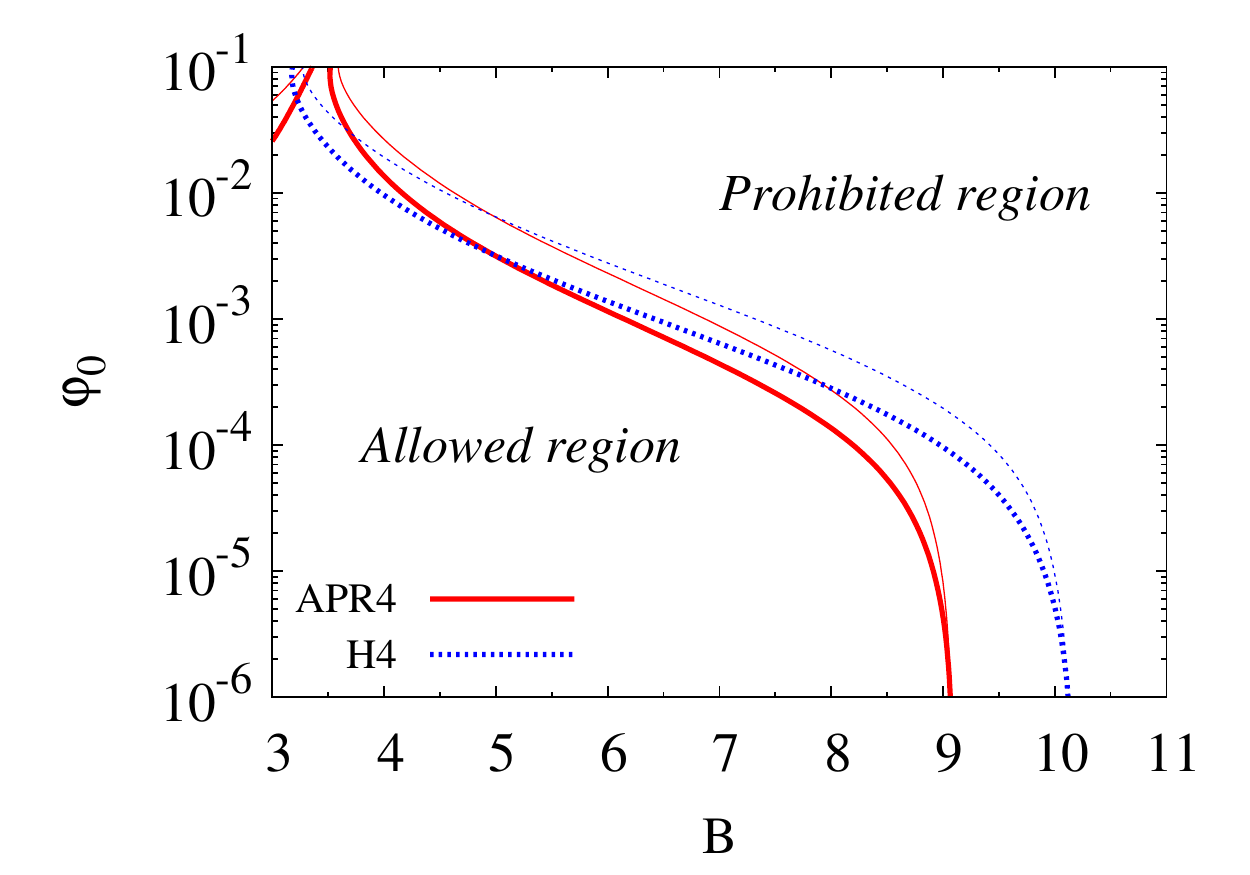}
\caption{The allowed region in the $B$-$\varphi_0$ plane 
derived from the constraint equation~(\ref{eq:cons1b}) setting the
pulsar mass to $1.46M_\odot$ for PSR\,J1738+0333. The thick and thin
solid curves show the result for $\alpha_r=0.05$ and 0.2,
respectively. At $B \approx 3.5$ for the APR4 EOS and $B \approx 3.2$
for the H4 EOS, the dipole radiation is suppressed because the relation
$M_\varphi/M_{\rm NS} \approx M_{\varphi,{\rm WD}}/M_{\rm WD}$ is
satisfied~\cite{PSRJ1738,DEF2}. Note that the constraint by the
Cassini spacecraft~\cite{Cassini} is written as $B\varphi_0^2\alt 5
\times 10^{-5}$ and is stronger than that imposed by the binary pulsar 
for $B \alt 5$.}
\label{fig:Constr}
\end{figure}

In the following we work at leading order, that is we neglect all
higher-order, nonlinear corrections in the luminosity (see
Refs.~\cite{DEF3,WZ89,MW} for more precise results). The
gravitational-wave luminosity from the tensor quadrupole moment in a
binary system in circular orbits is
\beq
{dE \over dt}\Big|_{\rm tensor\; quad}={32 \over 5}
\left({\mu \over m}\right)^2
\left({m \over a}\right)^5,
\label{tensquad}
\eeq
where $m$, $\mu$, and $a$ are the total mass, the reduced mass  
$M_{\rm WD}M_{\rm NS}/m$, and
the orbital separation, respectively.  Hereafter, we consider binaries
composed of a neutron star of mass $M_{\rm NS}$ and a white dwarf of
mass $M_{\rm WD}$. We derive the scalar-wave luminosity from the
scalar dipole moment integrating Eq.~(\ref{eq:JBD2a}). The relevant
term in the wave zone is $\varphi
\rightarrow \dot d_i n^i /r$ where $n^i$ is the unit spatial vector
pointing along the radial direction and $d_i$ is the scalar dipole
moment with magnitude 
\beqn
{a \over m}\left|M_{\rm WD}M_\varphi-M_{\rm NS}M_{\varphi,{\rm WD}}\right|
=a\mu \left|{M_\varphi \over M_{\rm NS}}
-{M_{\varphi,{\rm WD}} \over M_{\rm WD}}\right|. 
\eeqn 
Here, $M_{\varphi}$ and $M_{\varphi,{\rm WD}}$ are the scalar charges 
of the neutron star and white dwarfs, and $\dot d_i=(d/dt)d_i$. 
Substituting this dipole-moment contribution into the stress-energy
tensor of the scalar field, we find that the scalar-wave luminosity
from the scalar dipole moment in a neutron star-white dwarf binary in
a circular orbit is
\beqn
{dE \over dt}\Big|_{\rm scalar\;dip}&=&{1 \over 6}
\left({2 \over B} - {1 \over 2}\varphi_0^2\right)
\left({\mu \over m}\right)^2
\left({m \over a}\right)^4 \nonumber \\
&& \hskip 1cm \times
\left({M_{\varphi} \over M_{\rm NS}}-{M_{\varphi,{\rm WD}} \over M_{\rm WD}}\right)^2.
\label{scadip}
\eeqn
Assuming that $B\varphi_0^2 \ll 1$, we write $2/B-\varphi_0^2/2
\approx 2/B \approx \omega_0 \varphi_0^2$ where $\omega_0$ denotes the 
asymptotic value of $\omega$, which has to be $\agt 4 \times
10^4$~\cite{Cassini}.  Thus in the following, we neglect the term
$\varphi_0^2/2$ in Eq.~(\ref{scadip}). The ratio of the luminosities
(\ref{tensquad}) and (\ref{scadip}) is
\beqn
\alpha_r := {(dE/dt)_{\rm scalar\;dip} \over (dE/dt)_{\rm tensor\;quad}}
={5 \over 96B}\left({M_{\varphi} \over M_{\rm NS}}
-{M_{\varphi,{\rm WD}} \over M_{\rm WD}}\right)^2
\left({a \over m}\right).
\nonumber \\
\eeqn
If observations constrain $\alpha_r$ to a certain value, 
then, the following constraint on $M_{\varphi}$ holds
\beqn
M_{\varphi} < \left(\sqrt{{96B \alpha_r \over 5}} 
\left({m \over a}\right)^{1/2} +{M_{\varphi,{\rm WD}} \over M_{\rm WD}}
\right)
M_{\rm NS},
\label{eq:cons1a}
\eeqn
for $M_{\varphi}/M_{\rm NS} > M_{\varphi,{\rm WD}}/M_{\rm WD}$ 
and 
\beqn
M_{\varphi} > \left(-\sqrt{{96B \alpha_r \over 5}} 
\left({m \over a}\right)^{1/2} +{M_{\varphi,{\rm WD}} \over M_{\rm WD}}
\right)
M_{\rm NS},
\label{eq:cons1a2}
\eeqn
for $M_{\varphi}/M_{\rm NS} < M_{\varphi,{\rm WD}}/M_{\rm WD}$.  We
notice that for large values of $B \agt 4$, $M_{\varphi}/M_{\rm NS} >
M_{\varphi,{\rm WD}}/M_{\rm WD}$. 

Currently, the strongest constraint on the DEF scalar-tensor
theory~\cite{DEF1,DEF2,DEF3} is due to the observation of the white
dwarf-neutron star PSR\,J1738+0333 system~\cite{PSRJ1738}. For this
system, $M_{\rm NS}=1.46^{+0.06}_{-0.05}M_{\odot}$,
$m=1.65^{+0.07}_{-0.06}M_{\odot}$, and the orbital period is
0.35479\,days with $\approx 0$ eccentricity. These data imply
$\sqrt{m/a} = (1.19 \pm 0.02) \times 10^{-3}$. For this binary system,
the decrease rate of the orbital period is measured with $\approx
12\%$ error and agrees with the prediction of general relativity
within $\sim 7\%$ at the one--$\sigma$ level.  This would imply that
in this binary system the scalar-wave luminosity cannot exceed $\sim
5\%$ of the gravitational-wave luminosity, i.e., $\alpha_r \alt
0.05$. The same qualitative conclusion would apply for the
PSR\,J0348+0432 binary~\cite{PSRJ0348}, which contains a neutron star
with mass $\sim 2M_\odot$.  

The numerical calculation shows that $M_{\varphi,{\rm WD}}/M_{{\rm
WD}} \approx B\varphi_0/2$ for low-mass white dwarfs with $M_{{\rm
WD}}\alt 0.2M_\odot$. This relation is also expected from
Eq.~(\ref{eq:varphi0}) with $T \approx -\rho$ which holds in the
Newtonian limit.  Thus, we employ this relation in the following.

Then, for PSR\,J1738+0333, we can write Eq.~(\ref{eq:cons1a}) as
\beqn
M_{\varphi} &<& \Biggl[5.1 \times 10^{-3}M_{\odot} 
\left({\alpha_r \over 0.05}\right)^{1/2}
\left({B \over 9}\right)^{1/2} 
\left({\sqrt{m/a} \over 1.19 \times 10^{-3}}\right)
\nonumber \\
&&
+6.57M_\odot\left({B \over 9}\right)\varphi_0
\Biggr]
 \left({M_{\rm NS} \over 1.46M_{\odot}}\right). \label{eq:cons1b}
\eeqn
Equation~(\ref{eq:cons1a2}) is also written in the similar form. 
Using these constraint relations for PSR\,J1738+0333 , we can
determine the allowed regions in the parameter space $B$--$\varphi_0$
of the scalar-tensor model. We do it constructing spherical-star
configurations with $M_{\rm NS}=1.46M_\odot$ and different values of
$B$ and $\varphi_0$.
In Fig.~\ref{fig:Constr} we show those allowed regions for a spherical
neutron star of mass $1.46M_\odot$. We find that $B$ has to be smaller
than $\approx 9.0$ and 10.0 for APR4 and H4 EOSs irrespective of the
value of $\varphi_0$.  Therefore, $M_\varphi \ll M_{\rm NS}$ and the
PSR\,J1738+0333 binary pulsar is not scalarized at the separation at
which it has been observed.  The allowed regions vary if we take into
account the one-$\sigma$ error for the mass of the pulsar. For
example, if the mass of the pulsar were $\approx 1.40M_\odot$, the
constraint is less severe (allowed region is slightly wider), whereas
if it were $\approx 1.50M_\odot$, the constraint is more severe.

It is straightforward to derive a constraint similar to
Eq.~(\ref{eq:cons1b}) for the PSR\,J0348+432 binary pulsar. Also in
this case we find that $M_{\varphi}$ has to be much smaller than
$M_{\rm NS} \sim 2.0M_\odot$. As a consequence, also the
PSR\,J0348+432 binary pulsar is not scalarized at the binary
separation at which it is observed.  However, as it can be seen in
Fig.~\ref{fig:MS}, for APR4 EOS, the constraint (\ref{eq:cons1b}) is
not as strong as the one we obtain for PSR\,J1738+0333, because this
pulsar has a large mass, so the relativistic effects are in any case
too significant to induce the scalarization (we note that this is also
the case for relatively soft EOSs in which the radius of $1.35M_\odot$
neutron stars is 11\,--\,12\,km).  For H4 EOS, we find that the value
of $B$ has to be smaller than $\sim 9.0$ and in this case, neutron
stars are scalarized up to $M_{\rm NS} \sim 2M_\odot$.

Thus to summarize, because of the constraints coming from the
observations of PSR\,J1738+0333 and PSR\,J0348+0432, $B$ has to be
smaller than $\sim 9.0$, both for APR4 and H4 EOSs. We note that for
stiff EOSs in which the radius of neutron stars is large $\sim 15$\,km
and the maximum mass for spherical neutron stars is larger than
$2.5M_\odot$, the constraint imposed by the observation of
PSR\,J0348+0432 is quite severe. For example, for MS1 EOS~\cite{MS1}
in which the radius of $1.35M_\odot$ neutron stars is $\approx
14.5$\,km, $B$ has to be smaller than $\sim 8.8$.

Finally, because the PSR\,J1738+0333 binary pulsar is not scalarized, 
we can use Eq.~(\ref{eq:MF}) to rewrite Eq.~(\ref{eq:cons1a}) as 
\beqn
&&\varphi_0 < 5.1 \times 10^{-5} 
\left({\alpha_r \over 0.05}\right)^{1/2}
\left({B \over 9}\right)^{1/2} 
\left({M_{\rm NS} \over 1.46M_{\odot}}\right)
\nonumber \\
&&\hskip 6mm \times 
\left({F-BM_{\rm NS}/2 \over 100M_{\odot}}\right)^{-1}
\left({\sqrt{m/a} \over 1.19 \times 10^{-3}}\right). 
~~~~~~\label{eq:cons3}
\eeqn
The above equation implies that $\varphi_0$ is smaller than $\sim
10^{-5}$ and $\sim 10^{-4}$ for APR4 and H4 EOSs with $M_{\rm
NS}=1.46M_\odot$, $\alpha_r=0.05$, and $B=9.0$, because for this mass,
$F \sim 500M_{\odot}$ and $\sim 50M_{\odot}$, respectively.  Thus,
$\omega_0 \approx 2/(B\varphi_0^2)$ has to be larger than $\sim
2\times 10^9$ and $\sim 2\times 10^7$ for APR4 and H4, respectively,
if $B$ is as large as $\sim 9$\,--\,10. These constraints are much
stronger than those given in Ref.~\cite{Cassini}, as also found in
Ref.~\cite{PSRJ1738}.
 
In the previous section, we have found that for $M_{\rm
NS}=1.35M_{\odot}$, the condition for the onset of dynamical
scalarization during the inspiral with the APR4 EOS is relatively
weak, $B \agt 8.0$. By contrast, with the H4 EOS, the condition for
dynamical scalarization is rather limited as $B\agt 9.0$.  Combining
the constraints derived in this section, we obtain the following
conditions for the onset of dynamical scalarization during the
inspiral stage: for the APR4 EOS, $8 \alt B \alt 9$, while for the H4
EOS, we find only a very narrow window in the vicinity of $B \sim
9.0$. These analyses clearly illustrate that the EOS of neutron stars
is a key ingredient to determine the onset of dynamical scalarization
in the inspiral stage.

\subsection{Choice of scalar-tensor parameters}\label{sec:IVc}

\begin{figure*}[t]
\includegraphics[width=86mm,clip]{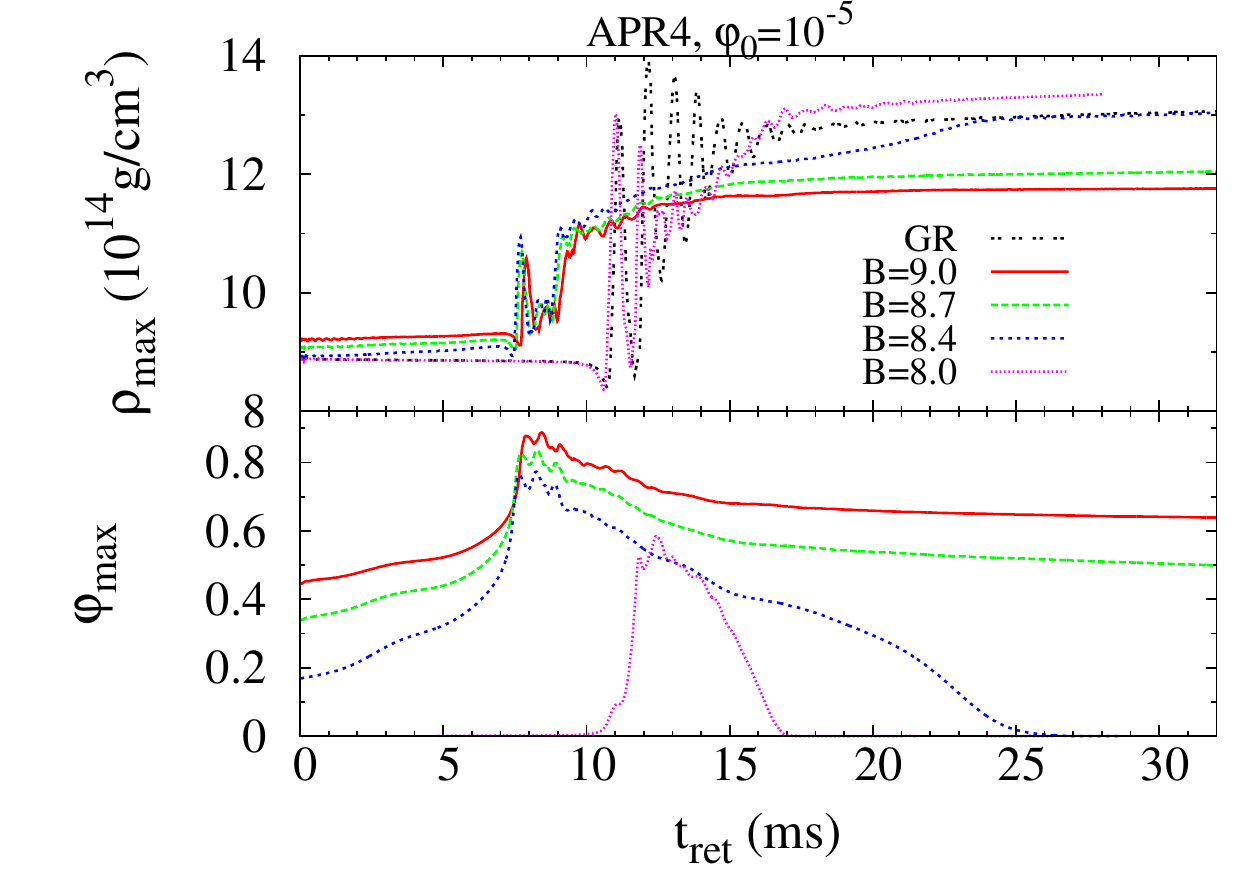}
\includegraphics[width=86mm,clip]{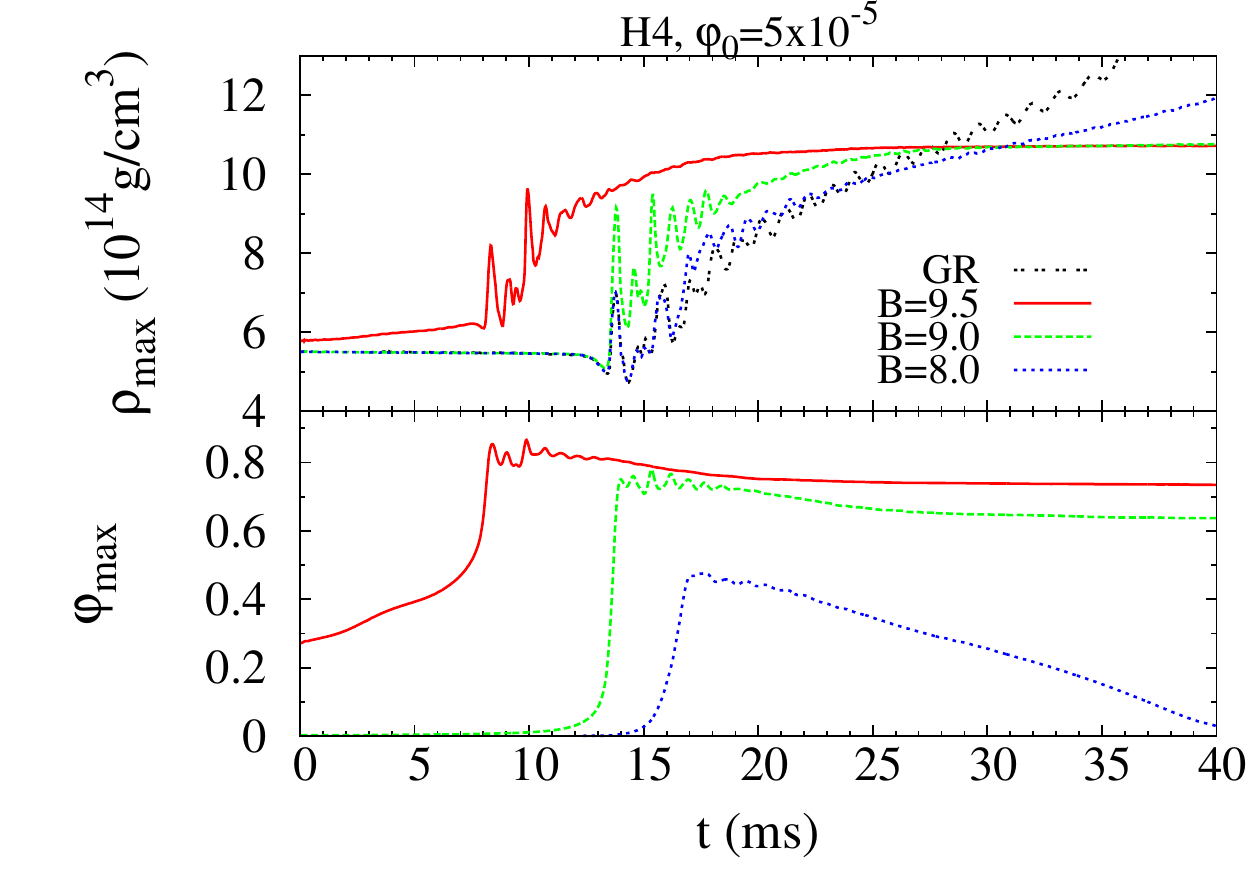}
\caption{Evolution of the maximum values of the rest-mass density 
and scalar field $\varphi$ for several models of $m=2.7M_{\odot}$ with
the APR4 EOS (left panel) and the H4 EOS (right panel). The merger
sets in at the time where the maximum density steeply increases. 
We note that for $B \geq 8.4$ with APR4 EOS and for 
$B=9.5$ with H4 EOS, the scalarization already occurred at $t=0$ 
(cf. Table~\ref{table:II}).}
\label{figrho}
\end{figure*}

Taking into account the constraints of the previous section, we employ
the following values of $B$ in the numerical simulations: $B=9.0$,
8.7, 8.4, 8.0, and 7.5 for the APR4 EOS, and $9.5$, 9.0, 8.5, and 8.0
for the H4 EOS.  The value of $B=9.5$ for the H4 EOS is not allowed by
the pulsar-timing observations as mentioned above.  However, we shall
investigate this case because we want to show that scalarization in
binary neutron stars occurs qualitatively in a universal manner
irrespective of the EOS employed. For small values of $B \alt 8.0$ for
APR4 and $\alt 9.0$ for H4, we do not expect dynamical scalarization
to occur during the inspiral stage.  However, the scalarization can
still occur in the {\em merger stage}.  This is why we employ such a
small value for $B$.

For a given value of $B$, the value of $\varphi_0$ is also constrained
(see Sec.~\ref{sec:IVA}). Taking into account the constraint given by
Eq.~(\ref{eq:cons3}), we choose $\varphi_0=10^{-5}$ for the APR4 EOS
and $5 \times 10^{-5}$ for the H4 EOS.  Note that the results
presented in this paper depend very weakly on the choice of
$\varphi_0$.

For the chosen values of $B$ with neutron-star mass $\approx
1.35M_\odot$, dynamical scalarization of neutron stars in a binary
system occurs for $a \alt 100M_{\odot} \approx 150\,{\rm km}$. For the
total mass of $2.7M_{\odot}$, this implies that dynamical
scalarization can occur only for $f \agt 100\,{\rm Hz}$ where $f$ is
the gravitational-wave frequency.  Therefore, due to the presence of
the strong constraints from the observations of
PSR\,J1738+0333~\cite{PSRJ1738} and PSR\,J0348+0432~\cite{PSRJ0348},
if neutron stars have canonical masses 1.3\,--\,$1.4M_\odot$, the
scalarization can take place only if the neutron star is in a compact
binary system.

In this paper we choose the initial value of the angular velocity as
$m\Omega=0.026$ for the APR4 EOS and 0.023 for the H4 EOS with
$m=2.7M_\odot$; the initial orbital period is 3.21\,ms and 3.63\,ms,
respectively; the initial separation is $a/m \approx
(m\Omega)^{-2/3}=11.4$ for the APR4 EOS and 12.4 for the H4 EOS; $a
\approx 31M_\odot$ for the APR4 EOS and $33M_\odot$ for the H4
EOS. Thus, for $B=9.0$, 8.7, and 8.4 with the APR4 EOS and for $B=9.5$
with the H4 EOS, for which $a < F$, dynamical scalarization has
already occurred at the initial separation (see Table~\ref{table:II}).
On the other hand, for $B\leq 8.0$ with the APR4 EOS and for $B\leq
9.0$ with the H4 EOS, dynamical scalarization has not yet occurred at
the initial separation because $a > F$.

\section{Numerical results}\label{sec:V}

\subsection{Characteristics of the merger process}

\begin{table}[t]
\caption{We list key quantities of our numerical simulations: 
EOS, the value of $B$, initial angular velocity in units of $m^{-1}$,
and total number of orbits. The total mass of the binary neutron stars
is $2.7 M_\odot$.  In the last column, we indicate when the
scalarization occurs. We consider that dynamical scalarization has
occurred when the value of $M_\varphi$ computed for a neutron-star in a
binary cannot be described by Eq.~(\ref{eq:MF}) (see Ref.~\cite{TSA13}
for details). }
{\begin{tabular}{cccccc} \hline 
~EOS~ & ~~~~$B$~~~~ & ~~$m\Omega$~~ & ~~Orbits~~ & ~~Scalarization\\ \hline 
APR4 & GR  & 0.026 & $\approx 5.0$ & --- \\ 
APR4 & 7.5 & 0.026 & $\approx 5.0$ & no scalarization \\
APR4 & 8.0 & 0.026 & $\approx 5.0$ & at merger\\
APR4 & 8.4 & 0.026 & $\approx 3.5$ & $m\Omega \approx 0.024$ \\ 
APR4 & 8.7 & 0.026 & $\approx 3.5$ & $m\Omega \approx 0.014$ \\ 
APR4 & 9.0 & 0.026 & $\approx 3.5$ & $m\Omega \approx 0.005$ \\ \hline
H4   & GR  & 0.023 & $\approx 5.0$ & --- \\ 
H4   & 8.0 & 0.023 & $\approx 5.0$ & after merger \\ 
H4   & 8.5 & 0.023 & $\approx 5.0$ & after merger \\ 
H4   & 9.0 & 0.023 & $\approx 5.0$ & at merger    \\ 
H4   & 9.5 & 0.023 & $\approx 3.0$ & $m\Omega \approx 0.017 $\\ 
\hline
\end{tabular}
}
\label{table:II}
\end{table}

In Fig.~\ref{figrho} we plot the maximum values of the neutron-star
density $\rho$ and scalar field $\varphi$ as functions of the time for
several values of $B$ and for the APR4 EOS (left panel) and the H4 EOS
(right panel) (see also Appendix B for a convergence study). For
comparison, we also plot the maximum density for the
general-relativistic case. Note that the merger sets in at the time
where the maximum density steeply increases. We observe the following
features of the merger process:
\begin{itemize}
\item For $B\leq 8.0$ with the APR4 EOS and for $B\leq 9.0$ 
with the H4 EOS, the maximum value of $\varphi$, $\varphi_{\rm max}$,
is always much smaller than unity before the onset of the merger. This
shows that for these models, the scalarization does not occur during
the inspiral stage as we expected in the analysis of
Sec.~\ref{sec:IV}.
\item Even for the initially weakly scalarized case (e.g., $B=8.4$ with 
the APR4 EOS), the scalar fields are amplified as the orbital 
separation decreases, signaling the occurrence of dynamical scalarization.
\item For the binary neutron stars that have scalarized, 
the duration of the inspiral stage is much shorter than that for the
nonscalarized case~\cite{B2013}.  For both EOSs, we find that starting
from the same initial frequency $m\Omega$, the inspiral stage of the
scalarized binaries is shorter than the one of binaries in general
relativity by 1\,--\,2 orbits. In the general-relativistic case the
inspiral stage lasts for $\approx 5$ orbits for both APR4 and H4 EOSs
(cf. Table~\ref{table:II}). Thus, the scalarization shortens the
inspiral stage by a significant fraction.  The reason for the
modification of the inspiral orbits for the scalarized case is that
the increase rate of the absolute value of the binding energy is {\em
decreased} by the scalarization effect.  We have also investigated
this effect using quasiequilibrium sequences of binary neutron stars
in Ref.~\cite{TSA13}. We notice that in the scalarized stage, the
orbital motion does not depend much on the values of $B$.
\begin{figure*}[t]
\includegraphics[width=86mm,clip]{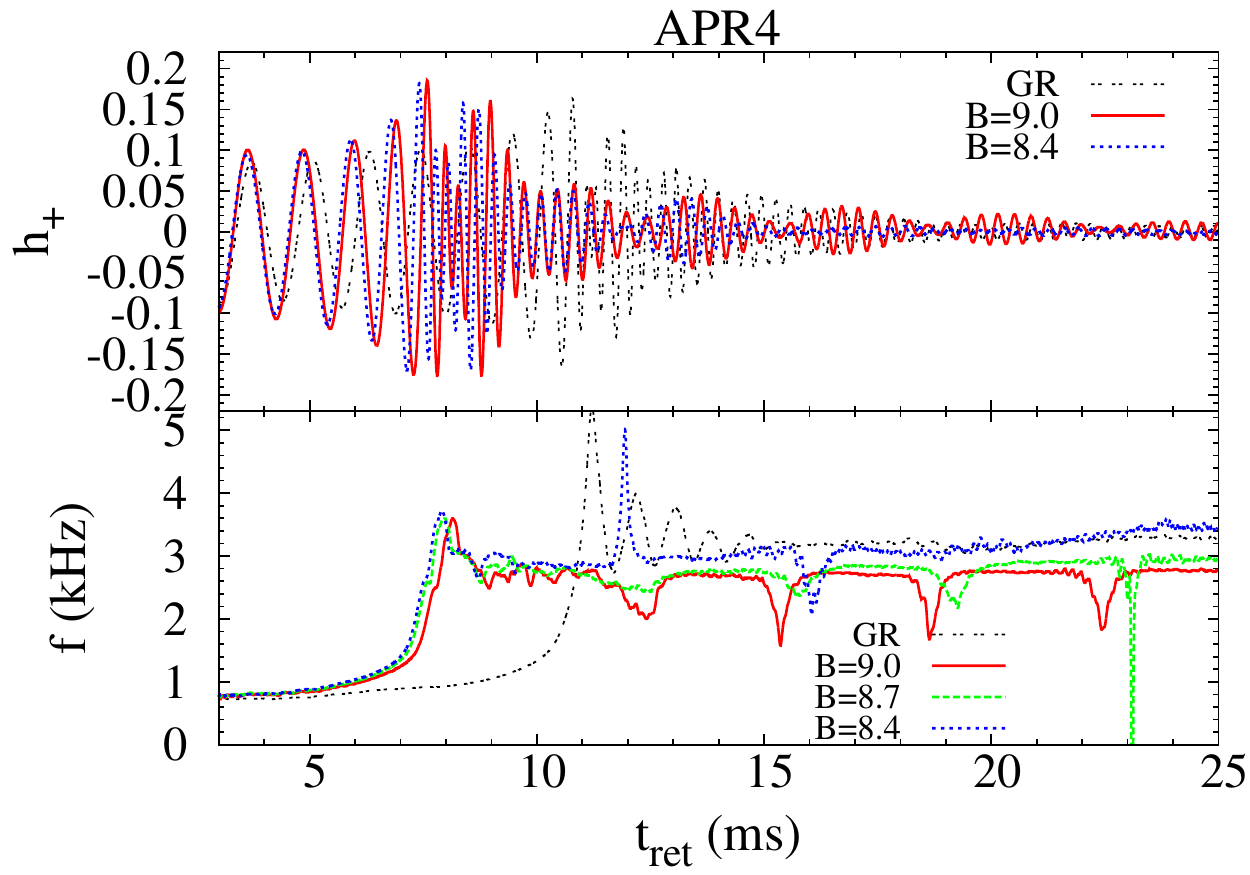}
\includegraphics[width=86mm,clip]{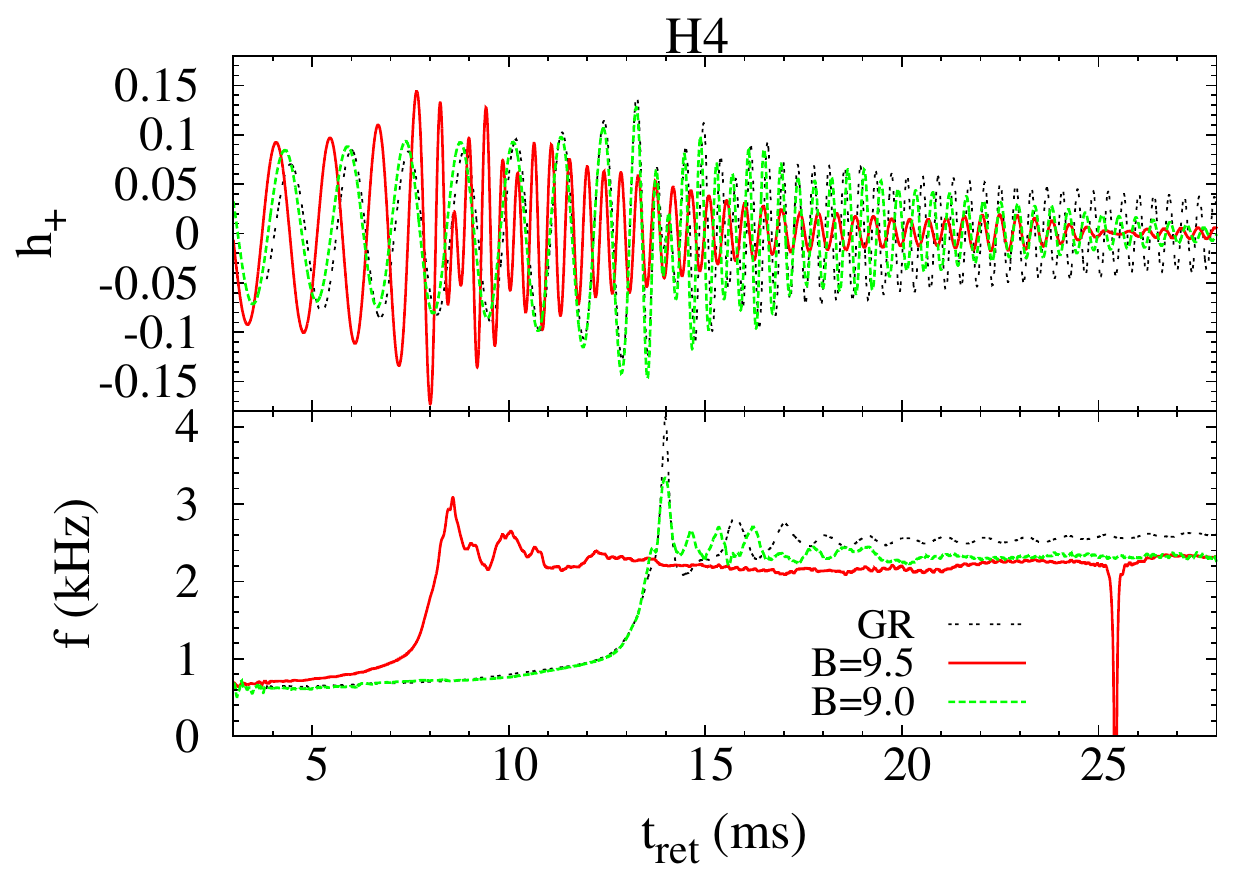}
\caption{We show plus-polarization gravitational waves 
observed along the axis perpendicular to the orbital plane and the
frequency of gravitational waves as functions of the retarded time for
$m=2.7M_{\odot}$ for the APR4 EOS (left panel) and the H4 EOS (right
panel). We note that when spikes occur in the frequency plots (at
$t_{\rm ret} \approx 23$\,ms for $B=8.7$ in the left panel and at
$t_{\rm ret} \approx 26$\,ms for $B=9.5$ in the right panel), the
amplitude of the gravitational waves is too low to accurately
determine the frequency.}
\label{fig:GW}
\end{figure*}
\item For the binary neutron stars that undergo dynamical scalarization 
(i.e., they have scalarized because of the presence of the companion),
the maximum density {\em increases} with the decrease of the orbital
separation.  This is in contrast with the general-relativistic case in
which the maximum density decreases with the decrease of the orbital
separation because of the tidal force exerted by the companion star.
The continuous increase of the maximum density in the scalarized
case is due to the fact that the amount of scalarization is enhanced with
the decrease of the orbital separation. 
\item For $B=8.0$ with the APR4 EOS and for $B=8.0$\,--\,9.0 
with the H4 EOS, the scalarization occurs {\em after} the onset of the
merger. (We note that for $B=7.5$ with the APR4 EOS, the scalarization
does not occur and hence the entire evolution is approximately the
same as that in the general-relativistic case.)  The maximum density
of the scalarized massive neutron star formed after the merger is
significantly different from that of nonscalarized or
general-relativistic cases.  This implies that the structure of the
scalarized remnant massive neutron star is also quite different from
the nonscalarized neutron star.
\item For relatively small values of $B$, the amplitude of the
scalar field of scalarized remnant massive neutron stars decreases
with time because their density increases, and eventually, the scalar
field approaches zero (see the curves for $B=8.4$ with the APR4 EOS
and $B=8.0$ with the H4 EOS). This is due to the fact that
relativistic effects become so strong during the evolution of the
remnant massive neutron star that the scalarization is turned off (see
Sec.~\ref{appendixB}).
\end{itemize} 

The reason of why the scalarization occurs after the onset of the merger, 
even for relatively small values of $B$, may be explained
using the analysis of Sec.~\ref{appendixB}. Indeed, we have found there 
that the scalarization is likely to occur for $(-BT)^{1/2}R
\rightarrow \pi/2$. Here, $R$ denotes the stellar radius. 
This implies that even for a small value of $BT$, the scalarization
can occur for a large value of $R$ or a large value of the
compactness, $\sqrt{-T}R \sim \sqrt{M/R}$ where $M$ is the mass of the
remnant massive neutron star. The compactness of the massive neutron
star of mass $\sim 2.6M_\odot$ is larger, by several 10\%, than the
compactness of a spherical neutron star of mass $1.35M_\odot$. Thus,
even for a small value of $B$ for which the scalarization cannot occur
for an isolated neutron star, the scalarization may occur when the
massive neutron star is formed as a remnant.

In addition, we find that the lifetime of remnant massive neutron
stars can be significantly changed by the scalarization.  For the H4
EOS with $m=2.7M_\odot$, the lifetime is several 10\,ms and hence
relatively short in general relativity~\cite{hotoke2013b}. This is
also the case when $B=8.0$ for the H4 EOS. In these cases, the angular
momentum of the massive neutron stars is primarily reduced by the
angular-momentum transport to the outer material, which is induced by
the torque exerted by the massive neutron star of an ellipsoidal
figure. After substantial spin-down, the massive neutron star
collapses to a black hole.  By contrast, in the presence of
scalarization (e.g., $B \agt 8.5$), the massive neutron star relaxes
to a quasistationary state of a smaller degree of nonaxisymmetry. This
seems to indicate that the scalar field contributes to the
redistribution of angular momentum. The scalarized massive neutron
stars seem to possess high angular momentum but the profile is not
significantly nonaxisymmetric.  Because these massive neutron stars
are hypermassive, they will collapse eventually to a black hole by
some dissipation or transport processes of angular momentum. However,
the lifetime seems to be much longer than that in general relativity.

\begin{figure*}[t]
\includegraphics[width=86mm,clip]{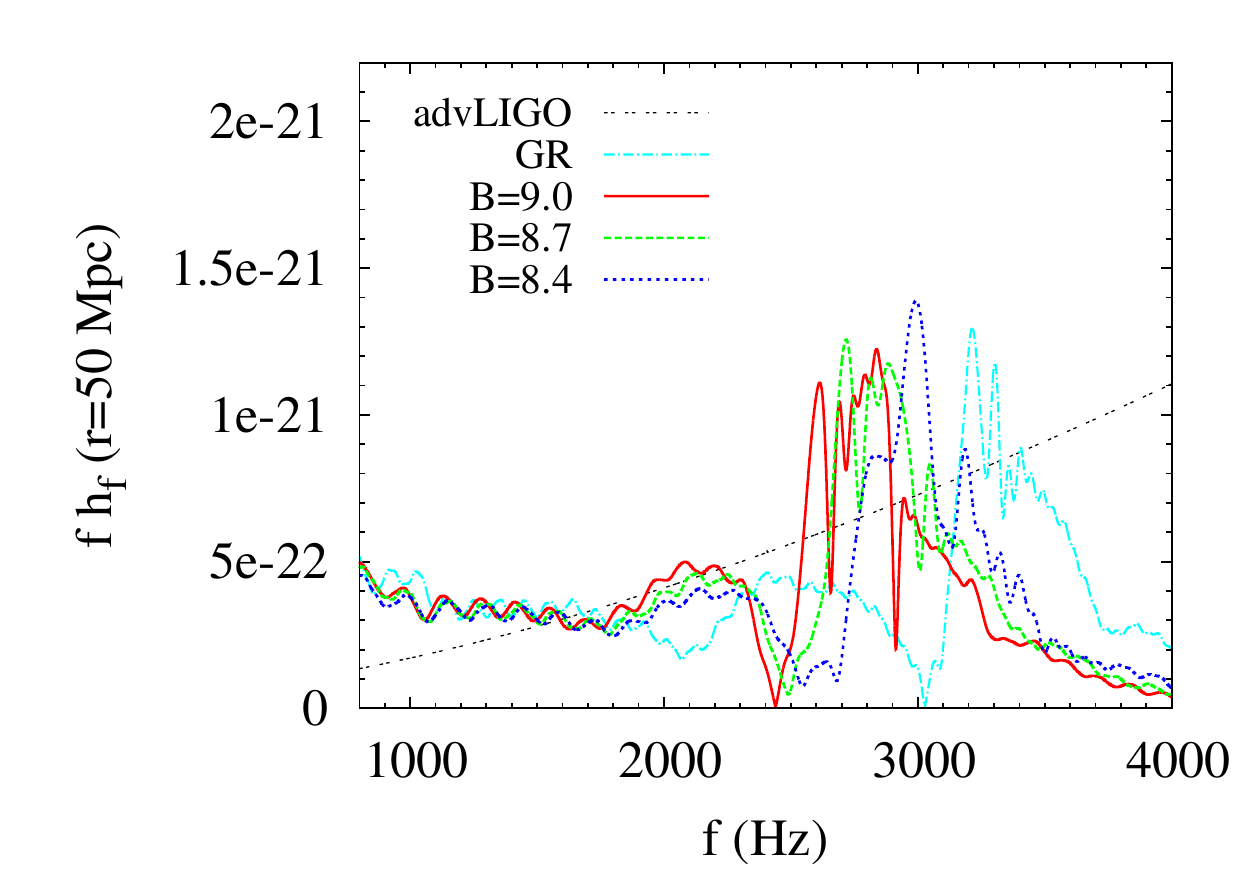}
\includegraphics[width=86mm,clip]{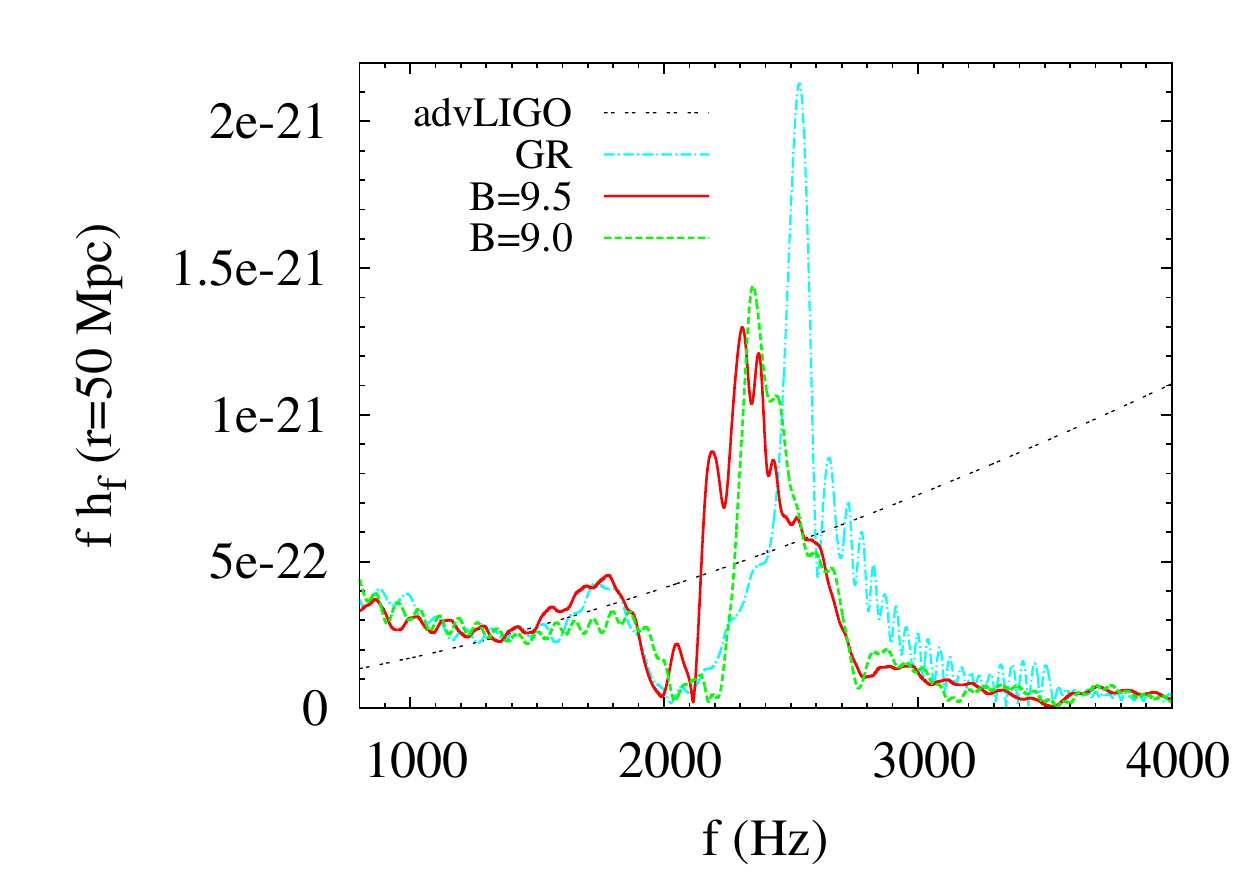}
\caption{The Fourier spectrum of gravitational waveforms 
for the APR4 EOS (left panel) and the H4 EOS (right panel).  We assume
that gravitational waves are observed along the axis perpendicular to
the orbital plane.  The black dot-dot curve is the noise spectrum
($\sqrt{f S_n(f)}$ with $S_n(f)$ being the noise power spectrum) of
the advanced LIGO with an optimistic configuration for the detection
of high-frequency gravitational waves (the so-called zero-detuned
high-power case: see
https://dcc.ligo.org/cgi-bin/DocDB/ShowDocument?docid=2974).  }
\label{fig:GWF}
\end{figure*}

Before ending this section, we briefly comment on the mass ejection
that could be a source of transient electromagnetic signals (e.g., see
Ref.~\cite{ejecta}).  Since the merger dynamics is modified by the
scalar field, we expect that the amount of ejected material is also
modified. For the APR4 EOS, the scalarized massive neutron stars
formed after the merger is less compact than that in general
relativity (see Fig.~\ref{figrho}).  In addition, the amplitude of the
quasiradial oscillations, which enhance angular-momentum transport,
are lower. Because of these effects, the total amount of ejected mass
is slightly decreased. Indeed, Ref.~\cite{hotoke2013} found that
compact massive neutron stars with high oscillation amplitude produce
larger mass ejection.  In general relativity, an equal-mass binary
with $m=2.7M_\odot$ ejects a mass of $\sim 7\times
10^{-3}M_\odot$~\cite{hotoke2013} while we find that for $B=9.0$, the
mass ejected is $\sim 5\times 10^{-3}M_\odot$.  Thus, the effect is
mild.  By contrast, the effect is significant for the H4 EOS. In this
case, the total amount of ejected mass is quite small in general
relativity $\sim 5\times 10^{-4}M_\odot$. However in the scalar-tensor
theory, it becomes $\sim 5\times 10^{-3}M_\odot$ for $B=9.5$ and $\sim
2\times 10^{-3}M_\odot$ for $B=9.0$. A possible reason of this finding
is that due to scalarization, the massive neutron star becomes more
compact and hence the effect of shock heating is enhanced and more
material is ejected.

\subsection{Gravitational-wave characteristics}

As a result of the modification of the dynamical motion induced by the
scalarization, gravitational waveforms are also modified.  We show in
Fig.~\ref{fig:GW} the gravitational waveforms and the corresponding
frequencies for several values of $B$ and for the APR4 EOS (left
panel) and the H4 EOS (right panel). We also show the Fourier spectrum
of these gravitational waves in Fig.~\ref{fig:GWF} at a distance of
$50 {\rm Mpc}$. As described above and also found in
Ref.~\cite{B2013}, the inspiral stage shortens when the neutron stars
are scalarized, e.g., typically the number of gravitational-wave
cycles in the scalarized case is smaller than in the
general-relativistic case by 2\,--\,4 cycles. We obtain this reduction
simulating a binary evolution that is not very long. The difference in
number of cycles between the scalarized and general-relativistic cases
would increase for much longer waveforms. Long, accurate evolutions
are beyond the scope of this paper. They will be investigated in the
future using also comparisons with post-Newtonian models. Thus, in the
following we focus on the merger waveforms.

The modification of the waveform emitted by a massive neutron star
formed after the merger is quite evident even for relatively small
values of $B \agt 8.0$ both for the APR4 and H4 EOSs. However, the way
in which the merger waveform is modified depends on the EOS. For the
APR4 EOS, we find that a scalarized remnant massive neutron star is
less compact than a massive neutron star in general relativity. As a
result, the frequency of quasiperiodic gravitational waves is
significantly (down to $\sim 0.5$\,kHz) decreased due to the
scalarization (see Fig.~\ref{fig:GWF}). In general relativity, the
peak frequency is 3.2\,--\,3.3\,kHz while for $B=9.0$, it is much
lower 2.6\,--\,2.8\,kHz. We also note that the spectrum around the
peak is rather wide for the large values of $B \sim 9.0$. This
reflects the fact that the frequency of quasiperiodic gravitational
waves varies with time.

For the H4 EOS, the scalarized massive neutron star formed 
after merger is more compact than the one in general relativity. 
However, the frequency of quasiperiodic gravitational waves does not 
become higher; rather, it becomes slightly lower due to the scalarization. 
This indicates that not only the compactness but also the presence of 
the high-amplitude scalar field plays an important role for determining the
oscillation-mode frequency.  For the H4 EOS, it is also remarkable that the
damping time scale of the wave amplitude for the scalarized case is
shorter than in general relativity. The reason for this is that the
ellipticity of the massive neutron star decreases in a shorter time
scale for the scalarized case.

It is worth to emphasize that these modifications are seen even in 
the case for which the scalarization does not occur during the inspiral
stage. For such cases, the inspiral signal is not modified and cannot be
used to constrain the scalar-tensor theory. For such a small value of
$B$, the effects of the scalar field cannot be
observed in standard neutron stars, as well, and in the next section 
we shall discuss some implications of these findings. 

In addition to gravitational waves, scalar waves~\footnote{Those
scalar waves should not be confused with the scalar mode of
gravitational waves in a scalar-tensor theory.} produced by the scalar
field $\varphi$ can carry away nonnegligible energy from the
system. However, we find that the energy emitted is a small fraction
of the total energy dissipated. For example, for the APR4 EOS with
$B=9.0$, we find that scalar waves are emitted in both the late
inspiral and merger stages.  Even for this case, the total energy
emitted in scalar waves is only $\sim 3\%$ of that emitted in
gravitational waves.  For the case of smaller values of $B$, this
fraction is smaller. The primary reason for this small contribution is
that for the equal-mass case, the dipole radiation is absent, and the
main contribution comes only from the monopole and quadrupole
radiation.  Therefore, the gravitational-wave emission primarily
determines the evolution of the binary system during the inspiral
stage, even if the scalarization occurs during the late inspiral
stage.

\section{Summary and discussion}\label{sec:VI}

In this paper we used numerical-relativity simulations to investigate
the late inspiral and merger dynamics and the gravitational-wave
emission of binary neutron stars in a scalar-tensor theory that admits
spontaneous scalarization~\cite{DEF1,DEF2,DEF3}. 

We confirmed, through several numerical-relativity simulations, what
was suggested in Ref.~\cite{B2013}, notably that if one or both
neutron stars are not initially spontaneously scalarized, they can be
scalarized dynamically during the late inspiral stage due to nonlinear
interactions of the scalar field configuration (see also
Ref.~\cite{TSA13} for more details). After the scalarization sets in,
the inspiral is accelerated and the total number of gravitational-wave
cycles is significantly decreased with respect to the
general-relativistic case. Given the mass of the neutron star, its EOS
and the constraints from binary pulsar observations, we determined for
which values of $B$ dynamical scalarization occurs. For example for
$M_{\rm NS} = 1.35 M_\odot$, we found for the APR4 EOS, $8 \alt B \alt
9$, while for the H4 EOS, we found only a very narrow window in the
vicinity of $B \sim 9.0$ where the neutron stars in a binary system
are dynamically scalarized before coalescing. These results imply that
even if the DEF scalar-tensor theory may give deviations to general
relativity that are not detected by observations of binary systems at
large separations, i.e., in the weak-field regime (pulsar timing
observations), nevertheless, the binary system may undergo dynamical
scalarization during the last stages of inspiral, i.e., in the
strong-field regime, and produce larger deviations to general
relativity which could be detected by ground-based gravitational-wave
detectors. Further studies, which make use of longer
numerical-relativity waveforms, analytical templates to model them,
and data-analysis techniques of the kind employed in
Ref.~\cite{delpozzoetal}, will address and assess the interesting
possibility of observing such deviations to general relativity with
ground-based detectors.

Furthermore, we found that the scalarization can occur even after the
onset of the merger. The reason is that the newly formed, massive
neutron star can have larger compactness, and hence, the scalarization
can occur even for small values of $B$ for which standard-mass neutron
stars cannot be scalarized in a binary system (see
Sec.~\ref{appendixB}). We also found that the subsequent evolution of
the remnant massive neutron star is quantitatively different from that
in general relativity. When the remnant massive neutron star is
scalarized, the compactness is different from that in general
relativity and the frequency of quasiperiodic gravitational waves is
modified. The modification depends on the EOS. For the APR4 EOS, the
remnant massive neutron stars are less compact and the frequency of
the quasiperiodic oscillations is in general lower. By contrast for
the H4 EOS, the remnant is only slightly more compact and the
frequency of the quasiperiodic oscillations is not significantly
modified. Furthermore, the scalarization seems to enhance the
redistribution of angular momentum. In fact, we found that for
scalarized massive neutron stars, which are in general
nonaxisymmetric, the time scale of the decrease of the ellipticity of
the massive neutron star is shorter than in general relativity. As a
consequence of this effect, the gravitational-wave amplitude decreases
with a shorter time scale, and in addition, the life time of the
massive neutron star is increased.

For the case the scalarization occurs only after the merger, the
inspiral signal is the same as that in general relativity and cannot
be used to constrain the scalar-tensor theory. Nevertheless, we found
that quasiperiodic gravitational waveforms from the scalarized,
massive neutron stars are different from those in the
general-relativistic case. References~\cite{BJ2012,hotoke2013b}
discussed the possibility that the EOS of neutron stars can be
constrained by observing the frequency of those quasiperiodic
gravitational waves emitted by remnant massive neutron stars. Assuming
that general relativity is correct, this method could be
useful. However, our results showed that if general relativity is
slightly violated, the method proposed in
Refs.~\cite{BJ2012,hotoke2013b} alone is not sufficient to extract the
EOS because the frequency of quasiperiodic gravitational waves emitted
by remnant massive neutron stars depends not only on EOS but also on
the degree of scalarization.

Nevertheless, the results found in this paper suggest a new way of
testing general relativity. When $B$ is such that spontaneous and
dynamical scalarization does not set in before merger or they are very
weak, then the inspiral signal is not modified significantly, and the
EOS can be determined from the inspiral stage by observing finite-size
effects in binary neutron
stars~\cite{lai94,flanagan08,read09b,hinderer10,damour12,baiotti,bernuzzi,hotoke2012,read13}. If
for those values of $B$, the merger signal is modified because the
newly, formed neutron star is sufficiently massive to be scalarized
and one finds that the characteristic frequency of quasiperiodic
gravitational waves agrees with the prediction of general relativity,
then one would conclude that general relativity is correct also in the
strong field regime. However, if the characteristic frequency does not
agree with the general-relativity prediction, then one would find that
general relativity is violated. The success of this test depends
crucially on the sensitivity of the gravitational-wave detectors at
frequencies between $400$\,Hz and $\sim 4$\,kHz, on the statistical
significance of the quasiperiodic oscillations in the merger waveform
and on the possibility of producing numerical-relativity waveforms in
scalar-tensor theory with systematic errors smaller than statistical
ones.

\acknowledgments

We thank Enrico Barausse and Gilles Esposito-Far\'ese for useful
discussions.  This work was supported by Grant-in-Aid for Scientific
Research (24244028), by Grant-in-Aid for Scientific Research on
Innovative Area (20105004), and HPCI Strategic Program of Japanese
MEXT. H.O. acknowledges financial support provided
under the European Union’s FP7 ERC Starting Grant
``The dynamics of black holes: testing the limits of Einstein's theory''
grant agreement no. DyBHo–256667.
A.B. acknowledges partial support from NSF Grant No. PHY-1208881
and NASA Grant NNX09AI81G. A.B. also thanks for hospitality in the
longterm workshop on {\it Gravitational Waves and Numerical
Relativity} held at the Yukawa Institute for Theoretical Physics,
Kyoto University in May and June 2013.

\appendix

\section{Numerical simulations of isolated, spherical neutron stars}
\label{appendixA}

\begin{figure}[t]
\begin{center}
\includegraphics[width=86mm,clip]{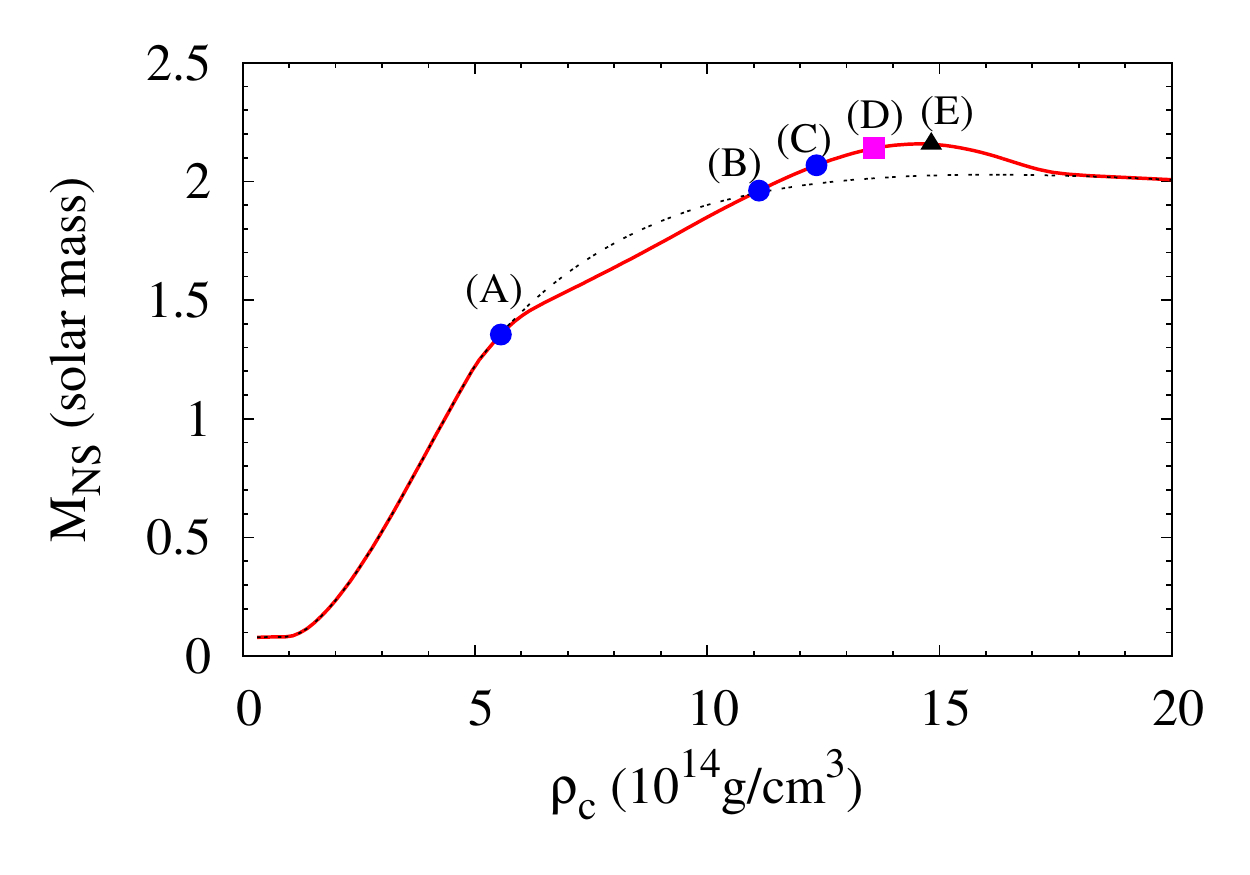}
\end{center}
\caption{The neutron-star mass as a function of the central density
$\rho_{\rm c}$ for spherical neutron stars with the H4 EOS with $B=10$
and $\varphi_0=3\times 10^{-3}$. The solid and dashed curves show the
relation in a scalar-tensor theory and in general relativity,
respectively.}
\label{fig:a1}
\end{figure}

To check the validity of our newly developed numerical-relativity code
for binary neutron stars in scalar-tensor theories, we perform
simulations of isolated, spherical neutron stars in scalar-tensor
theories.  We prepare spherical neutron stars using a piecewise
polytropic EOS (see, e.g., Ref.~\cite{hotoke2013} for details). We
perform many simulations varying the EOS and find that our conclusions
are essentially the same irrespective of the chosen EOS. For this
reason, here we focus on the results with the H4 EOS, which is a
rather stiff EOS, with the maximum mass of a spherical neutron star in
general relativity being $\approx 2.03M_{\odot}$ (see
Fig.~\ref{fig:a1}).

To show that our code works properly also in the case the coupling
between the scalar and tensor fields is strong, we choose $B=10$ and
$\varphi_0=3\times 10^{-3}$, even if those values are not realistic
because they were already excluded by the observation of neutron
star-white dwarf binaries (see Sec.~\ref{sec:IV}). For this choice of
the parameters, we plot in Fig.~\ref{fig:a1} the neutron-star (tensor)
mass as a function of the central density in general relativity and in
the scalar-tensor theory under investigation. As we see from
Fig.~\ref{fig:a1}, the mass in scalar-tensor theory starts differing
from the one in the general-relativistic case for $\rho_c \agt 5
\times 10^{14}\,{\rm g/cm^3}$ (or for $M\agt 1.35M_{\odot}$). We find
that this difference is a consequence of the fact the scalar field is
significantly excited, resulting in the modification of the density
profile of the neutron star.  When the central density is extremely
high, $\rho_c \agt 1.8\times 10^{15}\,{\rm g/cm^3}$, the neutron star
in general relativity and the scalar-tensor theory agrees with each
other approximately.  The reason is that $T(=T_a^{~a})$ becomes
positive in this density range, and thus the scalar mass becomes much
smaller than the neutron-star mass (see Sec.~\ref{appendixB}). We also
find that in the scalar-tensor theory, the maximum mass is $\approx
2.2M_{\odot}$, that is larger than in the general-relativistic
case. The fraction of increase depends strongly on the value of $B$ as
well as $\varphi_0$. All those properties are universal and
qualitatively independent on the EOS. 

We perform numerical simulations using five neutron stars, for which
the central density and neutron-star mass are plotted in Fig.~\ref{fig:a1}
(with labels (A)\,--\,(E)). The neutron stars (A)\,--\,(C) are expected
to be stable, while (D) and (E) could be unstable; in particular for
(E), it is reasonable to expect that it is unstable because the
central density is larger than that of the neutron star with the
maximum mass. 

\begin{figure}[t]
\begin{center}
\includegraphics[width=86mm,clip]{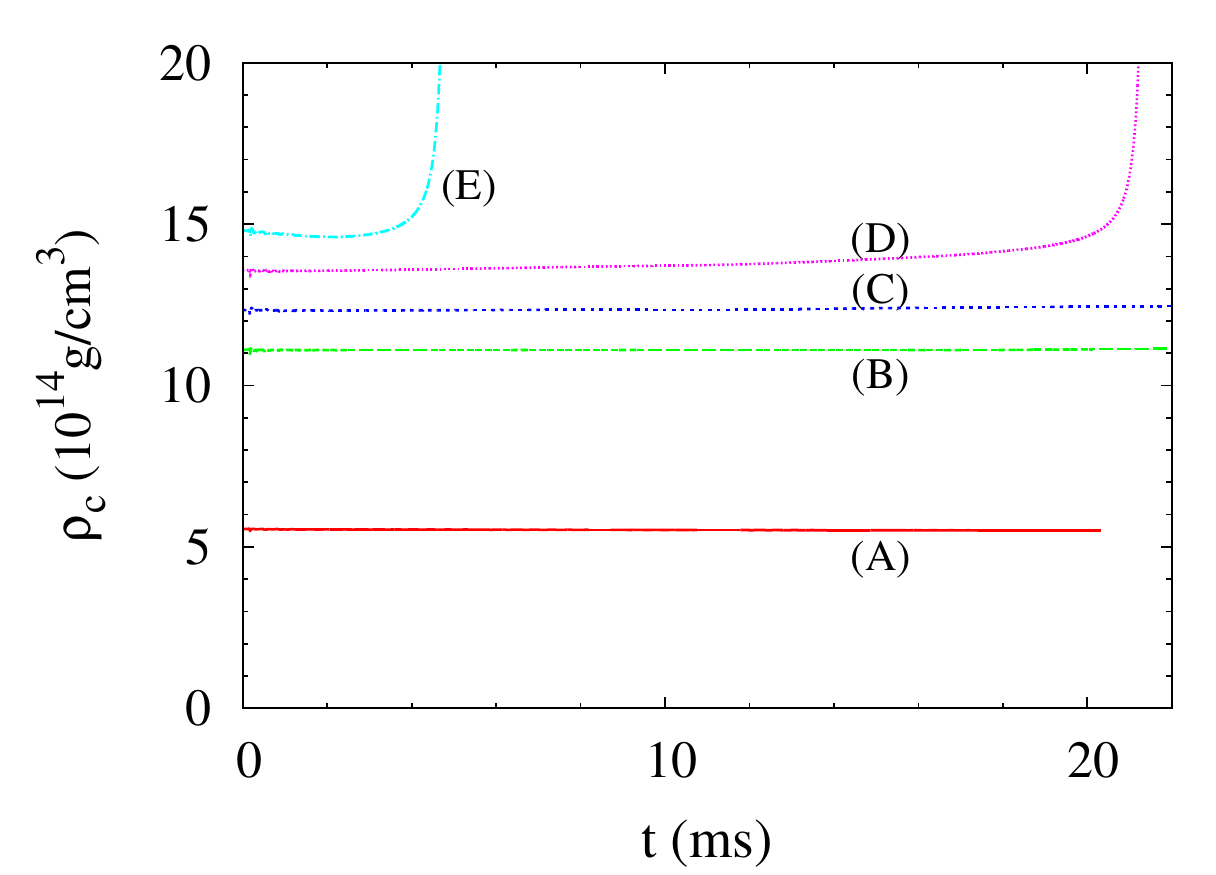}
\end{center}
\caption{Evolution of the central density $\rho_{\rm c}$ for five
spherical neutron stars. Curves from bottom to top show the results 
for (A) to (E), respectively. }
\label{fig:a2}
\end{figure}

We plot in Fig.~\ref{fig:a2} the evolution of the central density for
the five neutron stars. Note that the dynamical time scale of these
neutron stars defined by $\rho_{\rm c}^{-1/2}$ is shorter than
0.2\,ms, and hence, the simulations run for a time  much
longer than the dynamical time scale.  As expected, the neutron stars
(A)\,--\,(C) (having lower central density) are stable; the central
density (as well as the stellar structure) is unchanged in the
simulation time. By contrast, for (E), the star collapses to a black
hole in a short time scale.  Therefore, we conclude that it is
unstable against the radial oscillation. 

\begin{figure}[t]
\begin{center}
\includegraphics[width=86mm,clip]{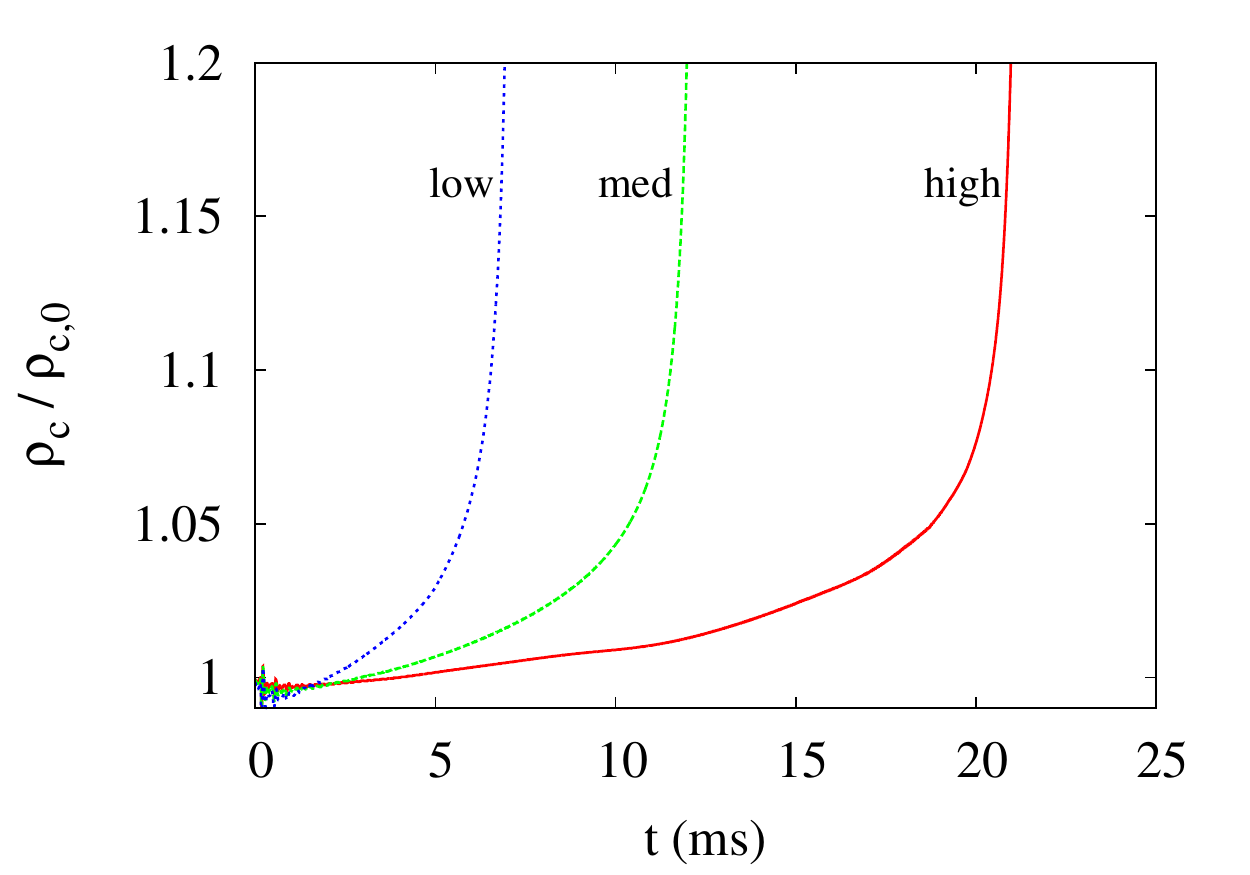}
\end{center}
\caption{Evolution of the central density for neutron star (D) 
(see Fig.~\ref{fig:a1}) with three different grid resolutions. The
dotted, dashed, and solid curves show the results for low, medium,
high resolutions, respectively, with the grid spacing, 0.369, 0.295,
and 0.236\,km.
}
\label{fig:a3}
\end{figure}

The stability of (D) is not very clear. In our simulation, this star
always collapses to a black hole in the time scale of 10
milliseconds. However, the lifetime depends strongly on the grid
resolution. In Fig.~\ref{fig:a3} we plot the evolution of the central
density for (D) with three different grid resolutions.  We find 
that the lifetime significantly increases as we improve the grid
resolution. Thus, we cannot draw a strong conclusion for
the stability to this neutron star. This finding is not surprising 
because the star (D) is located in the vicinity of the maximum mass
along the equilibrium sequence and thus it is likely that this star is
approximately equal to a marginally stable star. 

\begin{figure}[t]
\begin{center}
\includegraphics[width=86mm,clip]{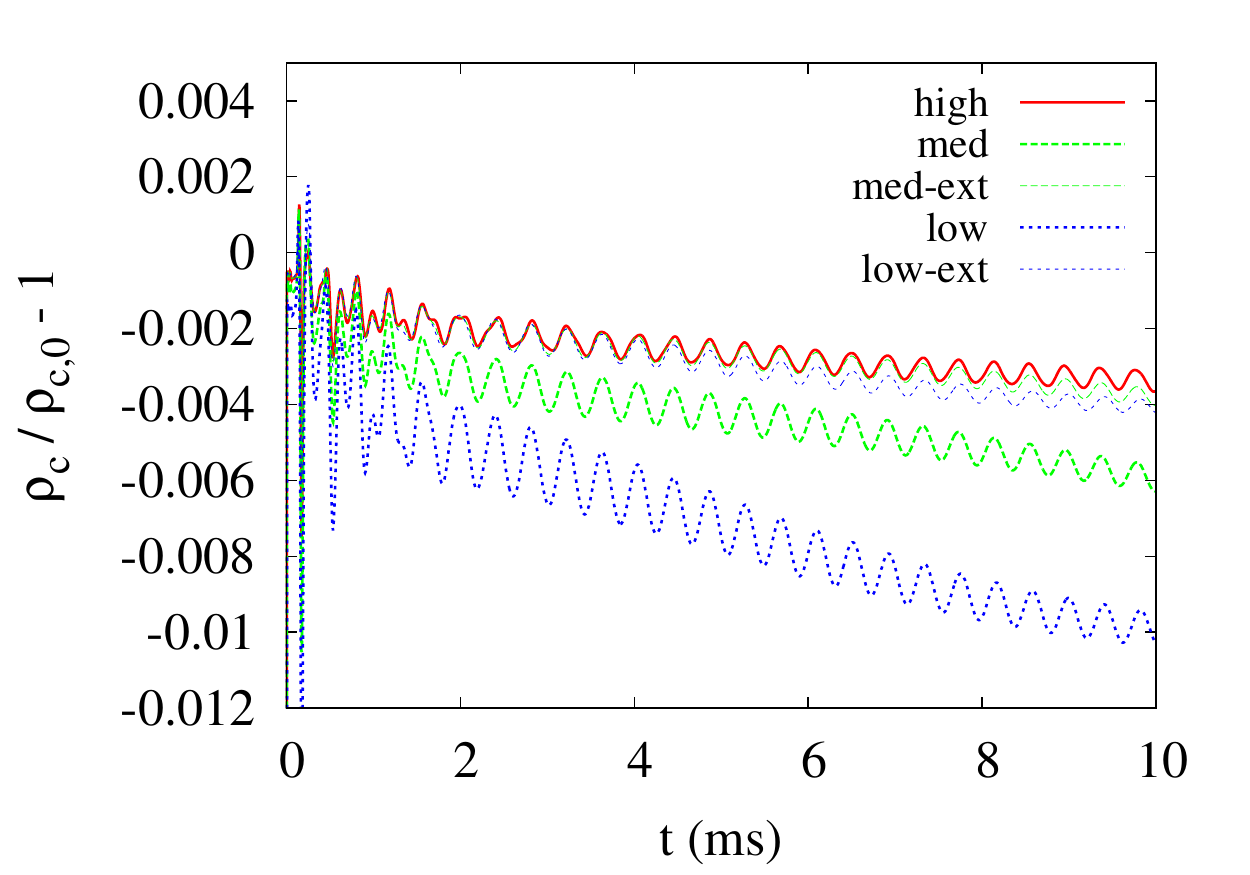}
\end{center}
\caption{
Evolution of the central density for neutron star (A) (see
Fig.~\ref{fig:a1}) with three different grid resolutions. Here,
$\rho_{\rm c}/\rho_{\rm c,0}-1$ is plotted with $\rho_{\rm c,0}$ being
the central density at $t=0$. The dotted, dashed, and solid curves
show the results for low, medium, and high resolutions, respectively,
with the grid spacing, 0.461, 0.369, and 0.295\,km. The thin curves
are plotted extrapolating the results of the low and medium resolution
runs under the assumption of second-order convergence.}
\label{fig:a4}
\end{figure}

By contrast, the convergence of the numerical results is achieved in a
much better manner for the evolution of stable neutron stars. In
Fig.~\ref{fig:a4} we plot the evolution of the central density for (A)
with three different grid resolutions. Note that for this star,
the mass is approximately $1.35M_\odot$, i.e., approximately equal to
the neutron-star mass considered in this paper.  We find that due to
the numerical error, the central density gradually decreases with
time, but with improving grid resolution, such numerical effects
become smaller. To find the order of convergence, we also plot
$(\rho_{\rm c}/\rho_{\rm c,0}-1)(\Delta x_{\rm high}/\Delta x)^2$
where $\Delta x$ is the grid spacing and $\Delta x_{\rm high}$ is
$\Delta x$ for the high resolution run. We show in Fig.~\ref{fig:a4}
that this quantity agrees approximately with $\rho_{\rm c}/\rho_{\rm
c,0}-1$ for the high-resolution run.  This implies that the error
convergences approximately at second order.

\begin{figure}[t]
\begin{center}
\includegraphics[width=86mm,clip]{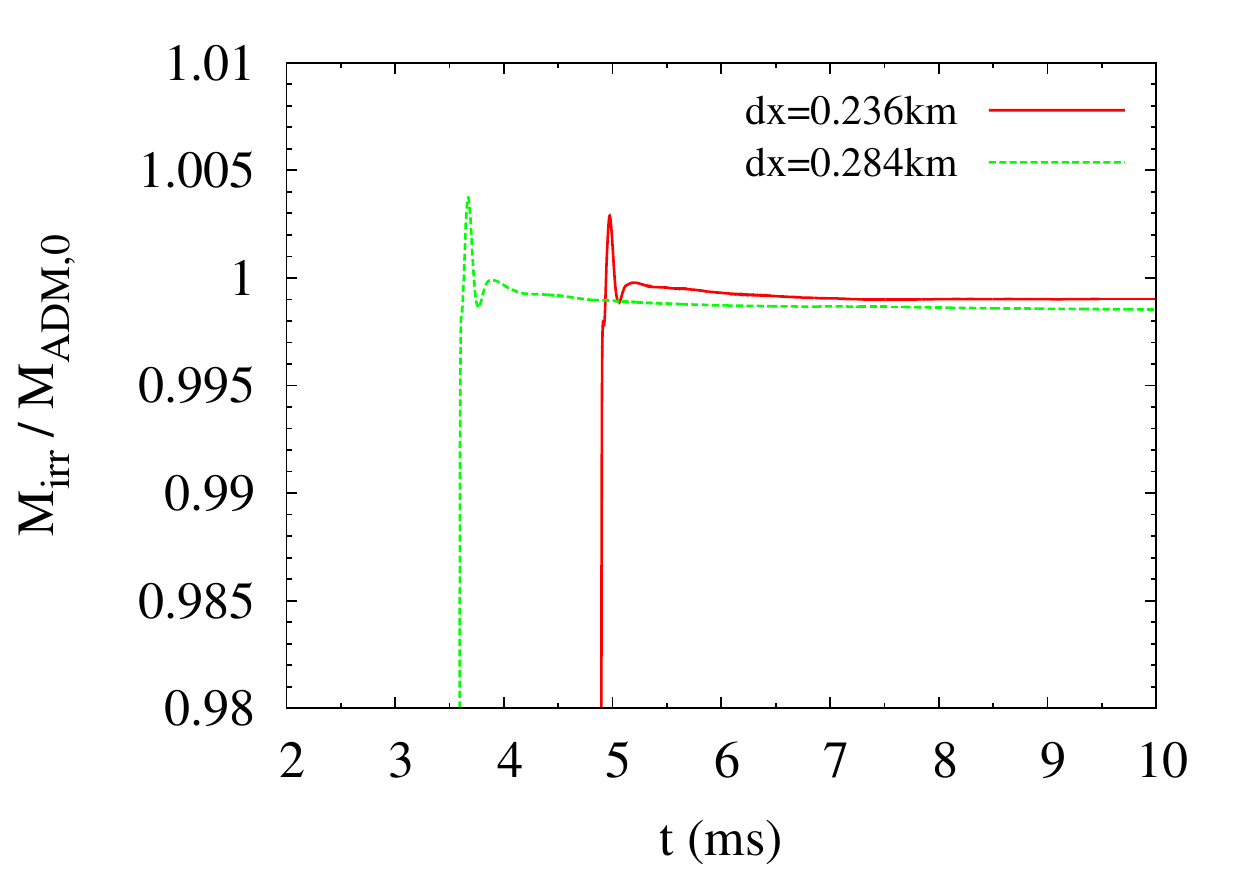}
\end{center}
\caption{Evolution of the irreducible mass defined by 
$\sqrt{A_{\rm AH}/16\pi}$ where $A_{\rm AH}$ is the area of the
apparent horizon for star (E) (see Fig.~\ref{fig:a1}) 
in units of the initial ADM mass.  }
\label{fig:a5}
\end{figure}

Finally, we show that we can accurately follow in our code the black
hole formation and evolution. In Fig.~\ref{fig:a5} we plot the
evolution of the irreducible mass defined by $\sqrt{A_{\rm AH}/16\pi}$
where $A_{\rm AH}$ is the area of the apparent horizon for neutron
star (E) in units of the initial ADM mass. Note that the ADM mass is
not equal to the tensor mass (neutron-star mass) and for this neutron
star, the initial Komar and tensor masses are 0.6\% and 0.3\% larger
than the initial ADM mass. After the formation of the black hole, the
scalar mass is lost, and hence, the mass of the black hole approaches
the initial ADM mass. We show in Fig.~\ref{fig:a5} that the final
black hole mass agrees with the initial ADM mass within 0.1\% for the
high-resolution run. Thus, the final mass does not agree with the
initial Komar mass nor the initial tensor mass. This indicates that
our code can follow the black hole accurately.

It is also worth to note that the irreducible mass {\em decreases}
with time in the early stage of the black hole evolution.  In general
relativity, this is not allowed. However, this is reasonable in the
present case because in the Jordan frame, the null energy
condition can be violated due to the presence of the scalar field, as
pointed out in Ref.~\cite{SST95}. 

\section{Convergence of numerical results}
\label{appendixC}

Here we want to discuss the convergence of the numerical-relativity
simulations. 

To check the convergence, we consider simulations for the APR4 EOS 
with $B=9.0$ and $\varphi_0=10^{-5}$ and for the H4 EOS with $B=9.5$
and $\varphi_0=5 \times 10^{-5}$. We perform runs using three grid
resolutions. For low, medium, and high resolutions, the stellar
diameter at the initial stage is covered approximately by 67, 80, and
100 grid points, respectively. 

\begin{figure}[t]
\begin{center}
\includegraphics[width=86mm,clip]{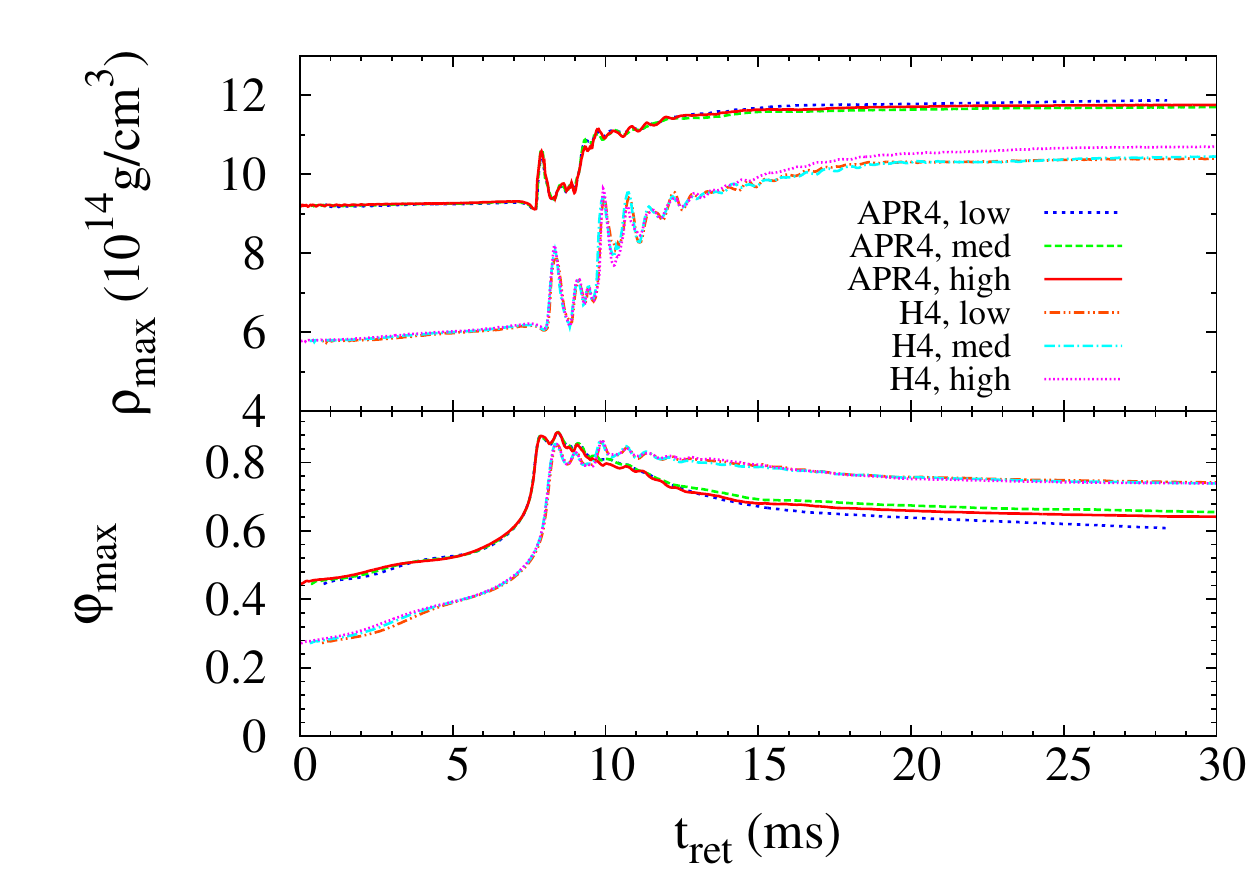}
\end{center}
\caption{The same as Fig.~\ref{figrho} for the APR4 EOS with $B=9.0$ and 
$\varphi_0=10^{-5}$ and for the H4 EOS with $B=9.5$ and $\varphi_0=5
\times 10^{-5}$ but for three different grid resolutions.  
We align the waveforms at the merger time by shifting the curves at
low and medium resolutions by approximately $+0.7$ and $+0.3$\,ms,
respectively.}
\label{fig:c1}
\end{figure}

In Fig.~\ref{fig:c1} we plot the evolution of the maximum density and
the maximum value of $\varphi$ for three grid resolutions.  As often
found in the simulations of inspiraling neutron stars, a lower grid
resolution always results in shorter merger time because of the larger
numerical dissipation. To align the merger time, we shift the curves
of low and medium resolutions approximately by $+0.7$ and $+0.3$\,ms,
respectively.  Although the inspiral duration is modified by the
numerical effect, Fig.~\ref{fig:c1} shows that the merger dynamics
depends only weakly on the grid resolution. Thus, for drawing the
conclusions in our paper, we can assume that we achieved convergence.

\begin{figure*}[t]
\includegraphics[width=86mm,clip]{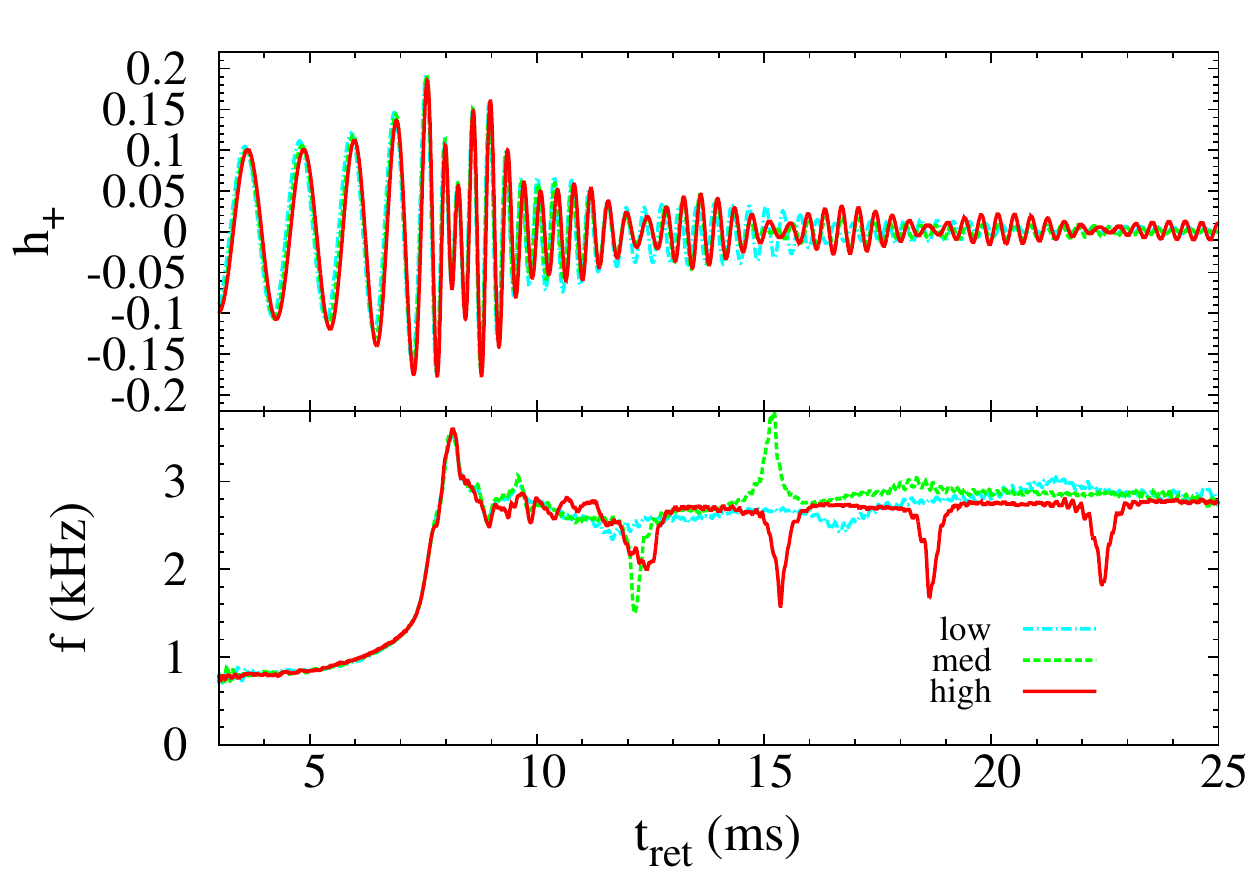}
\includegraphics[width=86mm,clip]{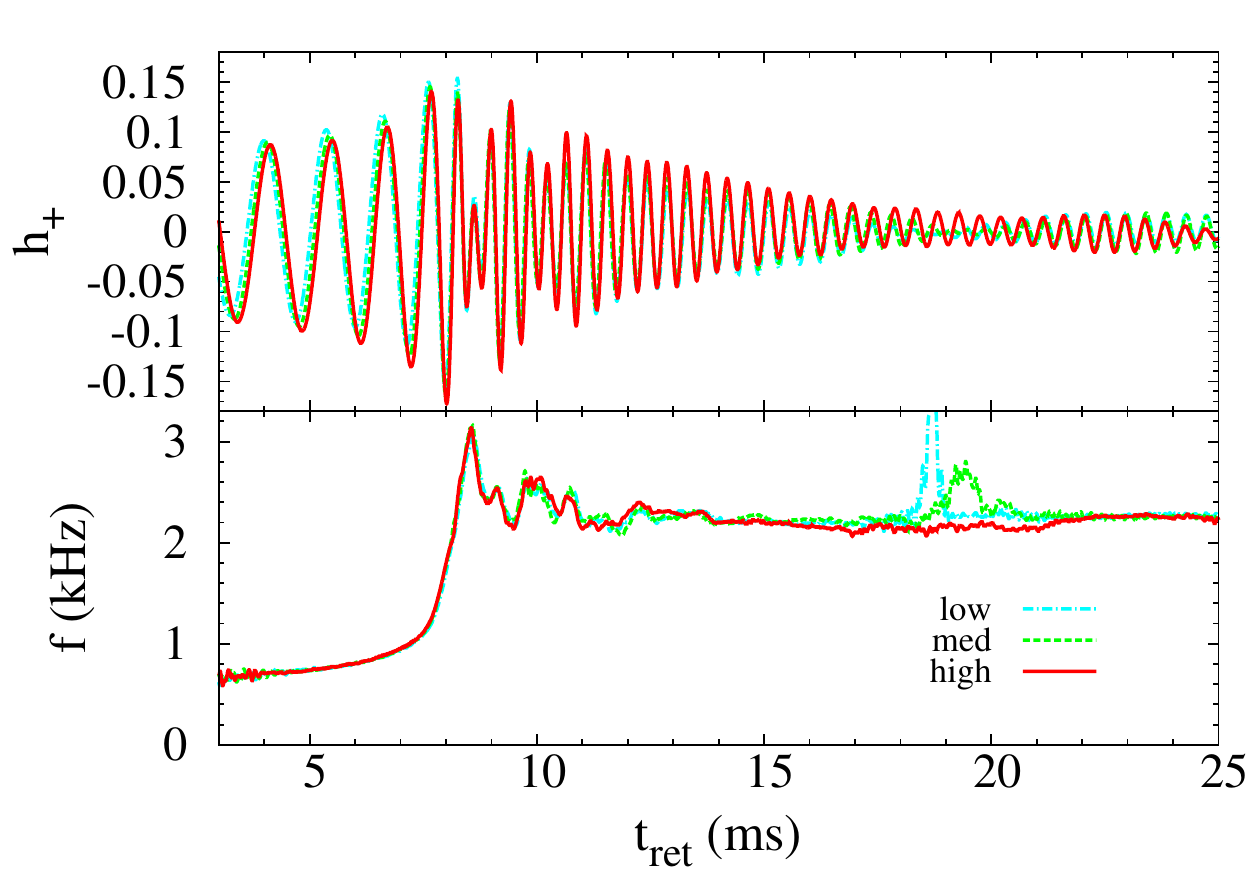}
\caption{The same as Fig.~\ref{fig:GW} for the APR4 EOS with $B=9.0$
and $\varphi_0=10^{-5}$ and for the H4 EOS with $B=9.5$ and
$\varphi_0=5 \times 10^{-5}$ but for three different grid resolutions.
We align the merger time by shifting the curves of low and medium resolution 
by approximately $+0.7$ and $+0.3$\,ms, respectively.}
\label{fig:c2}
\end{figure*}

We plot in Fig.~\ref{fig:c2} the gravitational waveform and the
corresponding frequency for the APR4 EOS with $B=9.0$ and for the H4
EOS with $B=9.5$.  Again, the time is shifted to align the waveforms
at merger. The waveforms computed with the three grid
resolutions agree qualitatively well, with the agreement being the
best with the H4 EOS.  For the early merger stage (i.e., in the first
$\sim 5$\,ms after the onset of the merger), the agreement is
quantitatively better independently on the EOSs. For the later merger
stage, the agreement becomes poorer, because the dynamics in the
merger stage depends strongly on the efficiency of shock heating for
which the convergence is achieved only at first order. Nevertheless,
the characteristic frequency of gravitational waves depends only
weakly on the grid resolution. We find that the disagreement is within
$\sim 0.1$\,kHz for the APR4 EOS and within $\sim 0.05$\,kHz or less
for the H4 EOS. Those differences are much smaller than the
differences from the general-relativity results. We find that the
convergence for the H4 EOS is much better than that for the APR4
EOS. The possible reason for this is that neutron stars with the H4
EOS are less compact and shock heating effects are weaker with this
EOS.

Finally, we extracted these gravitational waves at the finite radii
$\approx 200$\,--\,400\,km and we did not extrapolate the waveforms at
infinity because we expect that the numerical error due to the
extraction at finite radius is smaller than the one due to resolution
(see Refs.~\cite{hotoke2013c}).

\end{document}